\documentclass[lettersize,journal]{IEEEtran}
\usepackage{amsmath,amsfonts}
\usepackage{algorithmic}
\usepackage{algorithm}
\usepackage{array}
\usepackage[caption=false,font=normalsize,labelfont=sf,textfont=sf]{subfig}
\usepackage{textcomp}
\usepackage{stfloats}
\usepackage{url}
\usepackage{verbatim}
\usepackage{graphicx}
\usepackage{cite}
\usepackage{hyperref}

\usepackage[utf8]{inputenc}
\usepackage[normalem]{ulem} 
\usepackage{soul} 
\usepackage[table,x11names]{xcolor}
\definecolor{LightSkyBlue}{RGB}{135,206,250}
\definecolor{MediumPurple}{RGB}{147,112,219}
\definecolor{Violet}{RGB}{238,130,238}
\definecolor{SteelBlue}{RGB}{70,130,180}
\definecolor{YellowGreen}{RGB}{154,205,50}
\usepackage{longtable}

\usepackage{float}
\usepackage{makecell} 
\usepackage{colortbl} 
\usepackage{multirow}
\usepackage{comment}

\usepackage{pdflscape} 

\usepackage{amsmath,amssymb,amsfonts}
\usepackage{algorithmic}
\usepackage{graphicx}
\usepackage{textcomp}

\begin{document}

\title{Data Visualization for Improving Financial Literacy: A Systematic Review}

\author{
Meng Du$^{1,*}$,
Robert Amor$^{1}$,
Kwan-Liu Ma$^{2}$,
Burkhard C. Wünsche$^{1}$

\thanks{$^{1}$ School of Computer Science, The University of Auckland, Auckland, 1010, New Zealand. E-mail: mdu323@aucklanduni.ac.nz, r.amor@auckland.ac.nz, burkhard@cs.auckland.ac.nz}
\thanks{$^{2}$ Department of Computer Science, University of California, Davis, CA
95616, USA. E-mail: klma@ucdavis.edu}
\thanks{$^{*}$ Corresponding author. E-mail: mdu323@aucklanduni.ac.nz}
}

\markboth{Journal of \LaTeX\ Class Files,~Vol.~14, No.~8, August~2021}%
{Shell \MakeLowercase{\textit{et al.}}: A Sample Article Using IEEEtran.cls for IEEE Journals}

\IEEEpubid{0000--0000/00\$00.00~\copyright~2021 IEEE}

\maketitle

\begin{abstract}
Financial literacy empowers individuals to make informed and effective financial decisions, improving their overall financial well-being and security. However, for many people understanding financial concepts can be daunting and only half of US adults are considered financially literate. Data visualization simplifies these concepts, making them accessible and engaging for learners of all ages.
This systematic review analyzes 37 research papers exploring the use of data visualization and visual analytics in financial education and literacy enhancement. We classify these studies into five key areas: (1) the evolution of visualization use across time and space, (2) motivations for using visualization tools, (3) the financial topics addressed and instructional approaches used, (4) the types of tools and technologies applied, and (5) how the effectiveness of teaching interventions was evaluated. Furthermore, we identify research gaps and highlight opportunities for advancing financial literacy. Our findings
offer practical insights for educators and professionals to effectively utilize or design visual tools for financial literacy.
\end{abstract}


\section{INTRODUCTION}
\IEEEPARstart{F}{inancial} literacy, defined as ``a person’s ability to understand and make use of financial concepts'' \cite{servon2008consumer}, is crucial for everyday activities such as budgeting, risk assessment, and financial planning \cite{robb2009effect, kothakota2020use}. However, many individuals lack fundamental financial knowledge and numeracy skills, and only roughly half of US adults are considered financially literate \cite{tiaa2022pfin}. This leads to challenges like poor credit histories, insufficient retirement savings \cite{lusardi2007financial}, and financial insecurity. Traditional financial education often struggles to engage learners or address varying literacy levels, leaving many unable to make informed financial decisions. To tackle these issues, researchers have increasingly advocated for integrating data visualization into financial education \cite{sii2024investigating}. Visualization simplifies complex financial data, making it more accessible and engaging while enhancing trend identification and decision-making \cite{kothakota2020use, zhang2020research}.

While data visualization holds promise for transforming financial education, comprehensive reviews of its application remain scarce \cite{padilla2018decision, zabukovec2015impact}. Existing studies highlight its ability to simplify financial concepts and improve decision-making, yet gaps persist in understanding how to design and implement these tools effectively for diverse learner groups. Moreover, research in this area remains fragmented, lacking a unified synthesis to guide educators and tool developers. Addressing these gaps is critical for maximizing the potential of data visualization to foster financial literacy and decision-making skills.

In response, this systematic review synthesizes evidence on the use of data visualization and visual analytics in financial education. To provide a holistic perspective, we address five key research questions:

\textbf{RQ1:} How has data visualization for financial education evolved across temporal and spatial dimensions?

\textbf{RQ2:} What are the motivations for using data visualization to teach financial knowledge?

\textbf{RQ3:} What areas in financial education have researchers addressed, why, and how?

\textbf{RQ4:} What visualization tools and technologies are used in financial education and why?

\textbf{RQ5:} How were the teaching interventions evaluated and what results were obtained about their effectiveness?

Through this review, we aim to synthesize the current impact and capabilities of data visualization in financial education, providing practical guidance for educators, researchers, and practitioners in selecting effective visualization methods, promoting best practices in financial literacy education, and encouraging further research to enhance the use of data visualization in financial learning. Moreover, to the best of our knowledge, no systematic review exists in the field of visualization for financial education; this work fills that gap by offering a comprehensive analysis and valuable insights.
\IEEEpubidadjcol
\section{METHODOLOGY}
Our research methodology follows the systematic review process recommended by Booth et al.\cite{booth2021systematic}. First, we began by recognizing the need for a literature review, then formulated the research question and defined screening criteria. Next, we conducted searches and selected the publications for learning financial literacy by visualization from academic papers and online sources. Finally, we evaluated the chosen papers' relevance to our criteria and conducted data extraction. In this section, we describe the information collection procedure and the criteria we used to include literature related to visualization for learning financial knowledge.

\subsection{Search idioms and strategy}
To achieve comprehensive coverage across relevant research papers, broad search terms were first identified: ``Financial," ``Literacy," ``Education," and ``Data Visualization." Following the guidelines by Booth et al. \cite{booth2021systematic}, alternative spellings and synonyms were included, with Boolean operators ``OR" and ``AND" used to connect the primary terms effectively.

Alternative terms were identified to broaden the search scope. For ``Financial," alternatives included ``Finance" and ``Money." Substitutes for ``Literacy" included ``Knowledge," ``Understand," ``Understanding," ``Skill," and ``Skills," while ``Education" was expanded to ``Educate," ``Teaching," ``Teach," ``Learning," and ``Learn." During the search, it was noted that some papers did not use ``data visualization" specifically but included the broader term ``visualization." Consequently, ``data visualization" was substituted with commonly used terms in the visualization field, such as ``Visual," ``Visualize," ``Visualization," and ``Visual Analysis."

\begin{table}[h!]
\centering
\renewcommand{\arraystretch}{1.2}
\caption{Search string used for literature searching}
\resizebox{\columnwidth}{!}{
    \begin{tabular}{l}
    \hline
    (``Finance" OR ``Financial" OR ``Money") \\
    AND (``Literacy" OR ``Knowledge" OR ``Understand" OR ``Understanding" OR ``Skill" OR ``Skills") \\
    AND (``Visual" OR ``Visualize" OR ``Visualization" OR ``Visual Analysis") \\
    AND (``Learn" OR ``Learning" OR ``Teach" OR ``Teaching" OR ``Educate" OR ``Education") \\
    IN (Title OR Abstract OR Keywords) \\
    \hline
    \end{tabular}
}
\label{Search string}
\end{table}

Six major English-language databases were searched to ensure comprehensive and high-quality data collection: Scopus, Web of Science, IEEE Xplore, ACM Digital Library, Taylor \& Francis, and Wiley Online Library. Secondary searches of reference lists from included papers were conducted through Google Scholar. Table \ref{Search string} presents the search string, which is compatible across multiple databases.

\subsection{Eligibility criteria}
The inclusion criteria are as follows: (1) Must explain how the interventions aim to improve financial skills and understanding of financial concepts; (2) Must use data visualization as an intervention to improve financial skills and understanding; (3) Must have undergone a peer review to ensure its research quality and academic reliability. The exclusion criteria are as follows: (1) Must be a review paper, entire conference proceeding (rather than an individual paper), unpublished study, or a book; (2) Must not be written in English.

\subsection{Extraction process}
The data extraction process consists of three stages. First, the titles and abstracts were screened to exclude clearly irrelevant studies, e.g., about machine learning for stock market analysis using charts. Next, the full-text papers were examined to select relevant literature according to the above inclusion and exclusion criteria, focusing on aspects such as motivation, visualization intervention, and experimental evaluation. Any inconsistencies or disagreements were resolved through discussions among all authors. 

\subsection{Quality assessment}
In addition, a quality assessment was performed by three reviewers to assess the methodological soundness and reliability of the selected papers. This evaluation concentrated on three primary factors: (1) effective execution of the visualization intervention, (2) a research design that supports the extraction of pertinent information, and (3) verifying alignment with inclusion and exclusion criteria.

\begin{figure}[htb]
    \centering
    \includegraphics[width=\columnwidth, height=8.5cm]{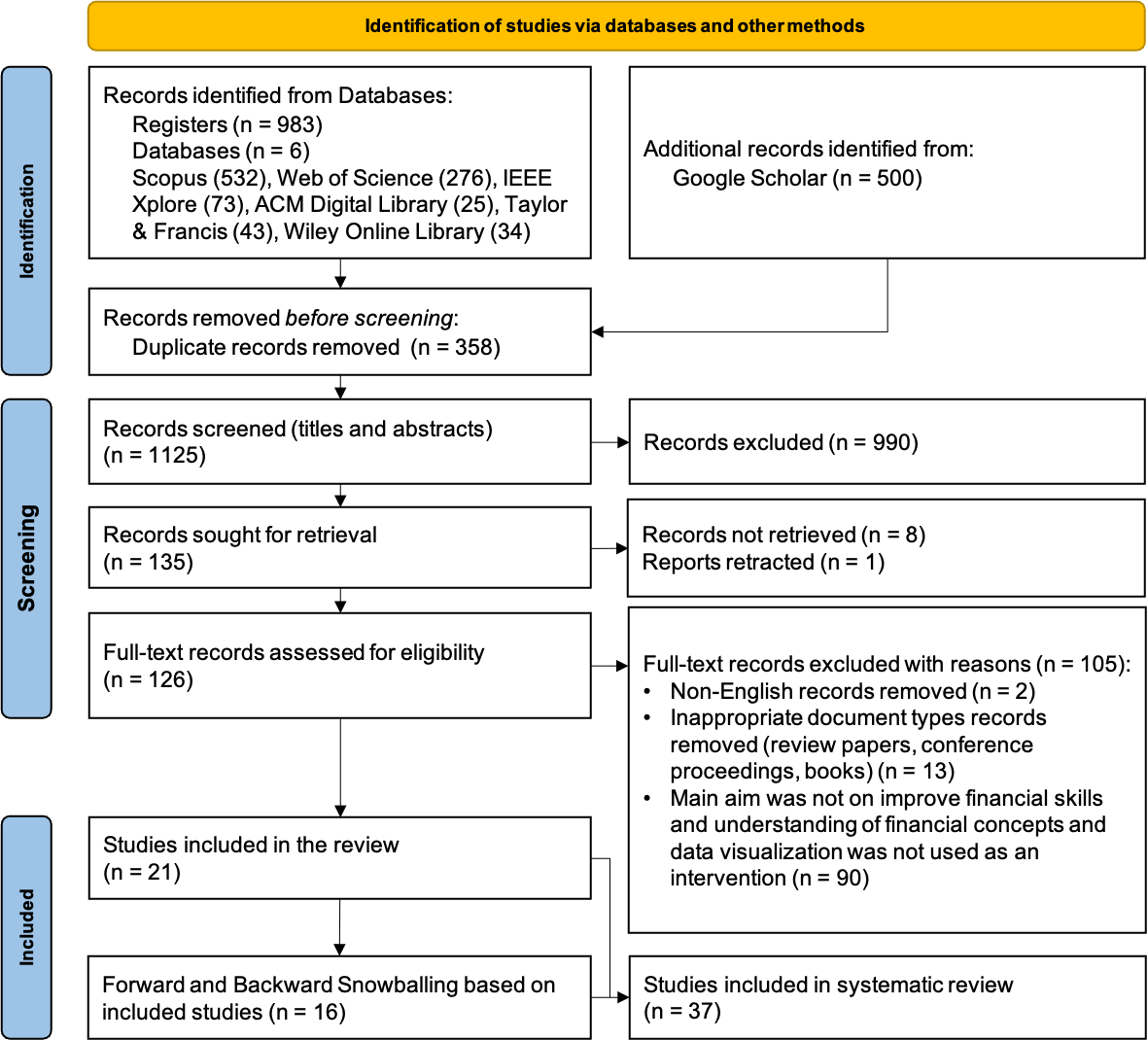}
    \caption{Systematic review PRISMA diagram, showing the selection and screening process.}
    \label{PRISMA}
\end{figure}

\subsection{Information extraction result}

A total of 983 papers were retrieved through the literature search: 532 from Scopus, 276 from Web of Science, 73 from IEEE Xplore, 25 from the ACM Digital Library, 43 from Taylor \& Francis, and 34 from Wiley Online Library. Additionally, 500 papers were identified through a secondary search on Google Scholar with the same search string. Given the large volume of results in Google Scholar, only the first 50 pages (10 papers per page) were reviewed, as they are ranked by relevance. After removing 358 duplicates, 1,125 unique papers remained. Following title and abstract screening based on the inclusion and exclusion criteria, 990 papers were excluded. Prior to full-text screening, 8 papers were found to be unavailable in full text, and 1 paper had been retracted. After the full-text review, 21 papers were included in the systematic review. A forward and backward snowballing search was then conducted on these papers, adding 17 more papers. In total, 37 papers were ultimately selected for the final review. Figure ~\ref{PRISMA} illustrates the study selection process. 
\section{RESULTS}
We reviewed 37 papers, categorizing the taxonomy of immersive visualization across five dimensions. Table \ref{main_table_5_year_trend} presents a summary of the key characteristics of the included studies, covering aspects such as motivation, financial knowledge, target audience, visualization type, and evaluation method. The boxes with different colors represent Motivation (\colorbox[HTML]{abdee6}{}), Financial Knowledge (\colorbox[HTML]{cbaacb}{}), Target Audience (\colorbox[HTML]{ffffb5}{}), Visualization Type (\colorbox[HTML]{ffccb6}{}), and Evaluation Method (\colorbox[HTML]{f3b0c3}{}). We also summarized five-year interval data and drew the trend lines at the bottom. In the following sections, we will discuss each research question and analyze the corresponding columns of this table.

\begin{figure*}[htbp]  
    \centering
    \makebox[\textwidth]{
        \includegraphics[width=\textwidth]{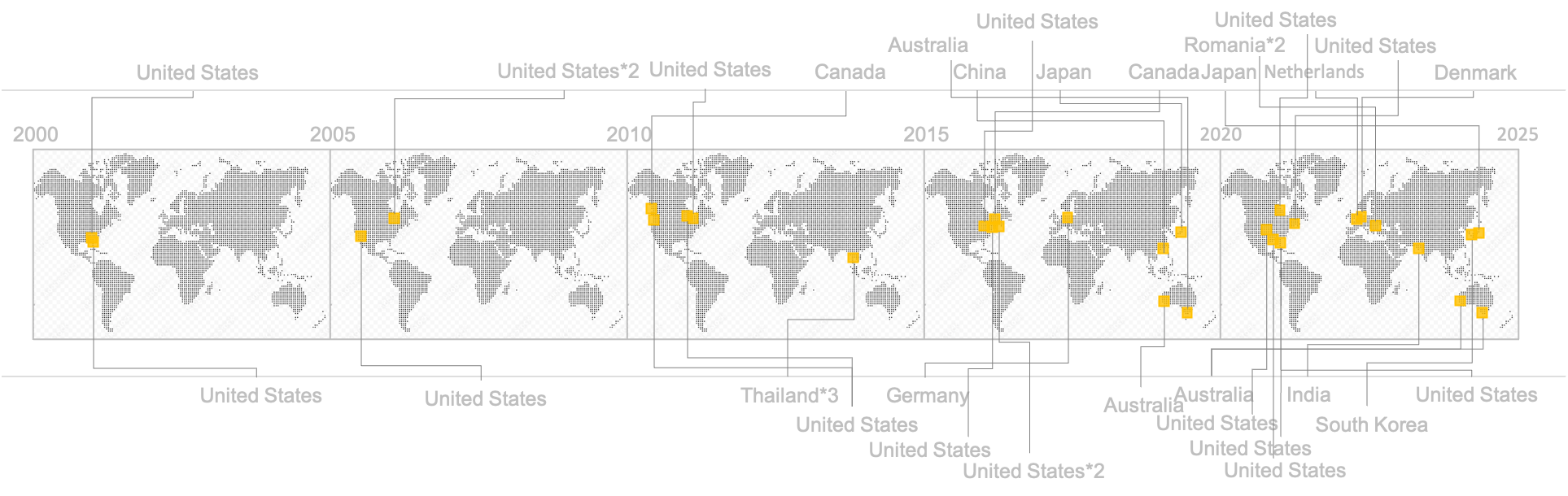}
    }
    \caption{Five-year spatial trends of selected publications. Yellow blocks mark the countries of first‐author affiliations, illustrating how research began in the US and gradually spread to a wider set of regions over each successive five‐year period.}
    \label{world_map}
\end{figure*}

\begin{table*}[htbp]
    \centering
    \caption{Eligible literature's characteristics.}
    \makebox[\textwidth][c]{\includegraphics[width=\textwidth,height=1\textheight,keepaspectratio]{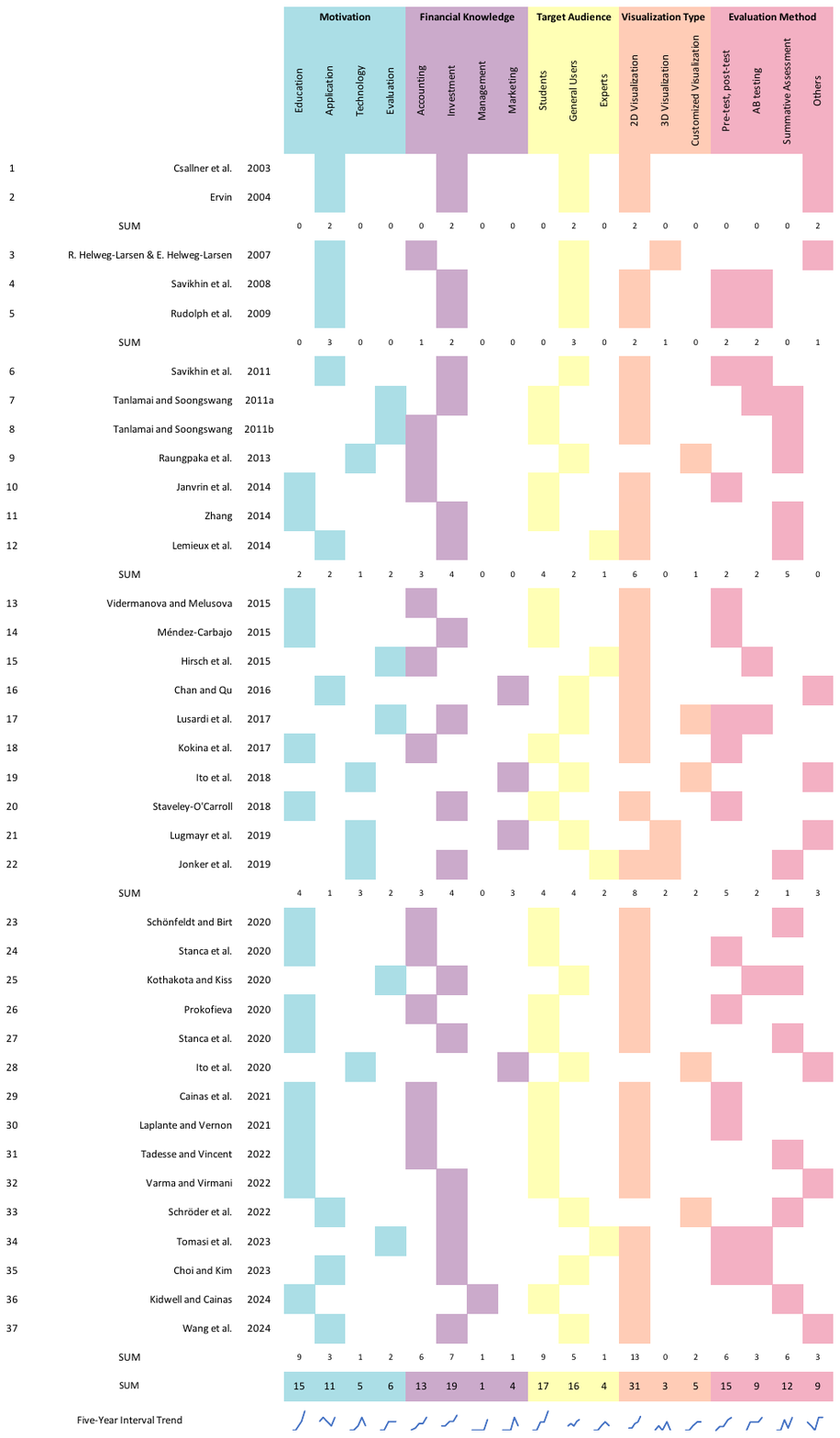}
    }
    \label{main_table_5_year_trend}
\end{table*}

\subsection{How has data visualization for financial education evolved across temporal and spatial dimensions?} 

\begin{figure}[htbp]
    \centering
    \includegraphics[width=0.9\columnwidth]{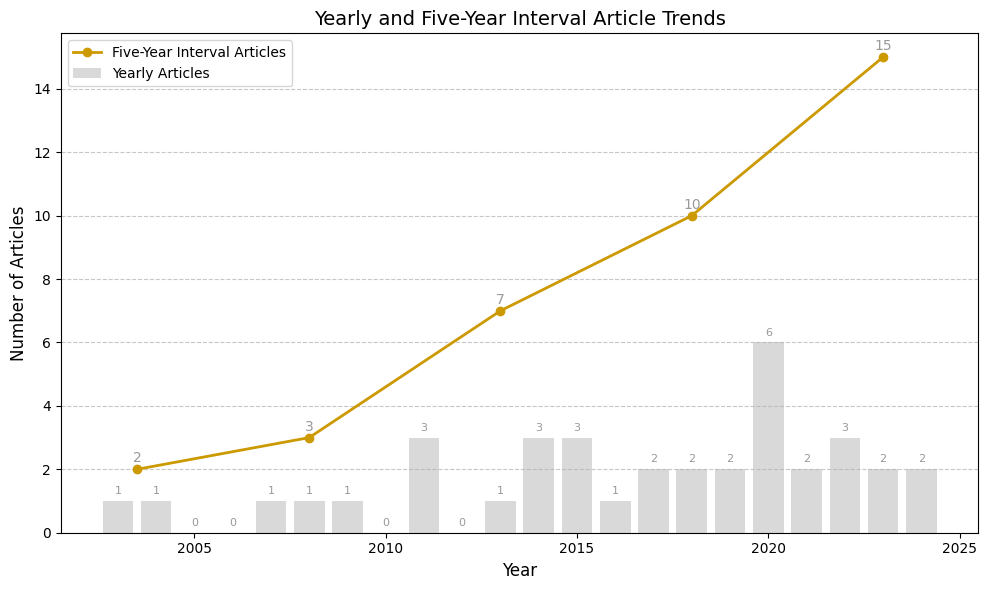}
    \caption{Yearly and five-year temporal trends of selected publications. Bars show annual publication counts and the line shows five-year totals, indicating a steady upward trend.}
    \label{bar_line_chart}
\end{figure}

This section highlights the spatiotemporal trends of publications. Figure~\ref{bar_line_chart} shows the bar chart of annual paper publications from 2003 to 2024. The earliest paper identified in our review that connects data visualization and financial literacy was published in 2003 \cite{csallner2003fundexplorer}, following the popularization of personal computers in 1984 \cite{azzam2013data}. By then, data visualization had evolved to allow direct user interaction \cite{chen2008brief}, entering the interactive phase in the 1990s. This suggests consistency between visualization development and its use in financial education. To analyze publication trends, the 20-year period from 2000 to 2025 was divided into five five-year intervals. The trend line chart shows a gradual rise in publications over these intervals. Five-year intervals were chosen to (1) reduce annual data fluctuations for clearer trend identification, especially for longer research spans, and (2) avoid years without publications, which can undermine analysis significance. The rising trend indicates increasing research interest in the field of financial literacy visualization, although, the most recent five-year period is influenced by a high number of publications in 2020. 

Figure~\ref{world_map} shows the country-of-origin distribution of published papers. Each map from left to right represents a five-year period. Yellow pixel blocks on the world map indicate regions with related publications, based on the first author's affiliation. Country names are listed above and below the map, arranged horizontally by time. If multiple papers are from the same region, they are marked with *number, indicating different authors. In the first two periods, only the United States published related papers\cite{csallner2003fundexplorer, ervin2004visualizing, helweg2007business, savikhin2008applied, rudolph2009finvis}. It was not until 2010-2014 that authors from Canada\cite{lemieux2014using} and Thailand\cite{tanlamai2011learning1,tanlamai2011learning2,raungpaka2013preliminary} published. In the last two periods, authors from more countries in Europe, Australia, and Asia contributed. The spatial distribution over these 25 years shows that authors from more countries are becoming interested in using visualization for financial education, although authors from the USA are still dominating this domain.


\subsection{What are the motivations for using data visualization to teach financial knowledge?} 

Table \ref{main_table_5_year_trend} highlights four key motivations in this field, which were identified through thematic grouping of the study objectives and research goals described in the selected papers. Education-driven research (15 papers) examines how visualization enhances teaching, learning outcomes, and course design. For example, three studies \cite{stanca2020impact1, cainas2021kat, tadesse2022combining} introduced innovative teaching methods using Excel and Tableau to improve student understanding and skills. Similarly, application-focused studies (11 papers) explore visualization's role in financial decision-making, literacy, and risk management, such as \cite{savikhin2008applied, rudolph2009finvis, savikhin2011experimental} applied visual tools to enhance users’ financial comprehension and decision-making. Technological advancements (5 papers) drive research on new visualization techniques, including 3D and VR, to expand financial analysis. Evaluation-focused studies (6 papers) assess visualization tools' effectiveness and impact on user comprehension and performance.

\begin{table}[htbp]
    \centering
    \renewcommand{\arraystretch}{1.5}
    \caption{Categories of Research Questions}
    \resizebox{\columnwidth}{!}{
    \begin{tabular}{>{\centering\arraybackslash}m{6.5cm} >{\centering\arraybackslash}m{1.5cm}}
        \hline
        \textbf{Paper} & \textbf{Question Type} \\ \hline
        \cite{csallner2003fundexplorer}
        \cite{ervin2004visualizing}
        \cite{helweg2007business}
        \cite{savikhin2008applied} 
        \cite{rudolph2009finvis}
        \cite{lemieux2014using} 
        \cite{cainas2021kat}
        \cite{savikhin2011experimental} 
        \cite{janvrin2014making} 
        \cite{zhang2014incorporating}
        \cite{vidermanova2015visualization}
        \cite{mendez2015visualizing}
        \cite{kokina2017role}
        \cite{ito2018text}
        \cite{staveley2018integrating}
        \cite{lugmayr2019financial} 
        \cite{jonker2019industry}
        \cite{schonfeldt2020ict} 
        \cite{prokofieva2020visualization}
        \cite{ito2020ginn}
        \cite{laplante2021incorporating}
        \cite{varma2022web}
        \cite{choi2023enhancing}
        \cite{kidwell2024tableau}
        \cite{wang2024visualization}
          & Application Method \\ \hline
        \cite{kothakota2020use}
        \cite{csallner2003fundexplorer}
        \cite{savikhin2008applied}
        \cite{rudolph2009finvis}
        \cite{lemieux2014using} 
        \cite{tanlamai2011learning1}
        \cite{tanlamai2011learning2}
        \cite{raungpaka2013preliminary} 
        \cite{stanca2020impact1}
        \cite{cainas2021kat} 
        \cite{savikhin2011experimental} 
        \cite{janvrin2014making}
        \cite{ito2018text}
        \cite{lugmayr2019financial}
        \cite{prokofieva2020visualization}
        \cite{ito2020ginn}
        \cite{wang2024visualization}
        \cite{stanca2020impact2}
        \cite{schroder2022pension}
        \cite{tomasi2023role}
        \cite{hirsch2015visualisation}
        \cite{lusardi2017visual} 
          & Impact Evaluation \\ \hline
        \cite{raungpaka2013preliminary}
        \cite{jonker2019industry}
        \cite{chan2016finavistory}
          & Design Strategy \\  \hline
        \cite{tadesse2022combining} &  \\ \hline
    \end{tabular}
    }
    \label{research questions}
\end{table}

Over the past 15 years, papers on educational motivations have steadily increased. In contrast, evaluation-focused research, which assesses the effectiveness and user impact of visualization tools, had no papers in the first two decades but has remained steady at about two per year since. Application and technological motivations, meanwhile, have shown fluctuating trends.

These motivations shape the direction of research and research questions in the reviewed papers. We grouped the research questions into two categories: explicit and unclear, as shown in Table \ref{research questions}. The explicit research questions fall into three types: ``application method” (how visualization is used), ``impact evaluation” (how well it works), and ``design strategy” (how to design effective visuals)—with 25, 22, and 3 papers in each category, respectively. Among them, only three papers \cite{kothakota2020use, tanlamai2011learning2, chan2016finavistory} stated their research questions, focusing on ``impact evaluation" and ``design strategy." For the remaining papers, while the research questions were not directly stated, they could be confidently inferred from the content. There was only one paper \cite{tadesse2022combining} for which we could not determine the research question, so it was classified as unclear.


By analyzing the alignment between research motivations and corresponding research questions, we observe clear patterns in how motivations shape inquiry. The type of motivation often determines the scope and direction of the research question. Educational motivations prompt inquiries into teaching practices, while technological motivations lead to methodological and design-focused exploration. Evaluation-driven papers consistently ask questions about effectiveness, reinforcing the link between purpose and inquiry.


The analysis of motivations and research questions reveals key gaps. Few papers focus on ``design strategy" as a research question, and even within this category, little research explores how to optimize designs to meet diverse needs \cite{raungpaka2013preliminary, jonker2019industry}. 

\subsection{What areas in financial education have researchers addressed, why, and how?} 

This section is divided into three parts. First, we analyze financial knowledge types, learning objectives, and target audiences, highlighting research focus areas and priorities. Next, we examine learning barriers and pedagogical methods, illustrating how the financial teaching was conducted. Finally, we assess feedback mechanisms to evaluate teaching effects on learners.

\subsubsection{Financial knowledge, learning objectives and target audience}

Table \ref{main_table_5_year_trend} categorizes financial knowledge and target audiences. Most studies apply visualization to investment (19 papers) and accounting (13 papers), both gaining significant attention over 25 years. In contrast, marketing (4 papers) and management (1 paper) receive less focus. Students are the primary audience (17 papers), with steady growth, while general users (16 papers) and experts (4 papers) show fluctuations.

We further analyzed the types, topics, and levels of learning objectives, as shown in Table \ref{knowledge_learning_obj_target}. Type defines the learning mode (e.g., skill-building vs. conceptual understanding), based on the stated instructional goals and descriptions in each paper. For example, if the objective involved mastering a software tool or applying a financial formula, we classified it as skill-building; if it emphasized understanding theoretical or structural concepts, it was categorized as conceptual understanding. Topic specifies targeted sub-goals, clarifying what learners aim to achieve within that mode. This distinction captures both the method and intent of each objective.


Regarding learning objective types, studies related to teaching finance courses at universities predominantly focus on ``learning to use tools" (11 papers), as seen in \cite{schonfeldt2020ict, cainas2021kat, tadesse2022combining}. However, the majority of papers emphasize ``learning to understand concepts" (26 papers), such as \cite{rudolph2009finvis, lusardi2017visual, schroder2022pension}.

\begin{table*}[htbp]
    \centering
    \renewcommand{\arraystretch}{1.5}
    \caption{Categories of Learning Objectives}
    \resizebox{\textwidth}{!}{
    \begin{tabular}[t]{>{\centering\arraybackslash}m{4.5cm} >{\raggedright\arraybackslash}p{14.5cm}}
        \hline
        \textbf{Learning Objective Type} & \multicolumn{1}{c}{\textbf{Paper}}
        \\ \hline
        Learn to use tools &  
         \cite{cainas2021kat}
         \cite{tadesse2022combining} 
         \cite{janvrin2014making} 
         \cite{zhang2014incorporating} 
         \cite{mendez2015visualizing} 
         \cite{kokina2017role} 
         \cite{staveley2018integrating} 
         \cite{schonfeldt2020ict} 
         \cite{laplante2021incorporating} 
         \cite{varma2022web} 
         \cite{kidwell2024tableau}
          \\
        Learn to understand concepts & 
         \cite{kothakota2020use} 
         \cite{csallner2003fundexplorer} 
         \cite{ervin2004visualizing} 
         \cite{helweg2007business} 
         \cite{savikhin2008applied} 
         \cite{rudolph2009finvis} 
         \cite{lemieux2014using} 
         \cite{tanlamai2011learning1} 
         \cite{tanlamai2011learning2} 
         \cite{raungpaka2013preliminary}
         \cite{stanca2020impact1} 
         \cite{savikhin2011experimental}
         \cite{vidermanova2015visualization} 
         \cite{ito2018text} 
         \cite{lugmayr2019financial}
         \cite{jonker2019industry} 
         \cite{prokofieva2020visualization} 
         \cite{ito2020ginn} 
         \cite{choi2023enhancing}
         \cite{wang2024visualization}
         \cite{stanca2020impact2} 
         \cite{schroder2022pension} 
         \cite{tomasi2023role}
         \cite{hirsch2015visualisation} 
         \cite{lusardi2017visual} 
         \cite{chan2016finavistory} 
        \\ \hline
        \textbf{Learning Objective Topic} & \multicolumn{1}{c}{\textbf{Paper}}
        \\ \hline
        Developing business accounting skills & 
         \cite{zhang2014incorporating}
         \cite{kokina2017role} 
         \cite{staveley2018integrating} 
         \cite{lugmayr2019financial}
         \cite{schonfeldt2020ict}
         \cite{laplante2021incorporating} 
         \cite{varma2022web} 
         \cite{kidwell2024tableau}\\
        Interpreting visualization & 
         \cite{kothakota2020use} 
         \cite{csallner2003fundexplorer} 
         \cite{ervin2004visualizing} 
         \cite{helweg2007business} 
         \cite{rudolph2009finvis} 
         \cite{lemieux2014using} 
         \cite{tanlamai2011learning1} 
         \cite{tanlamai2011learning2} 
         \cite{raungpaka2013preliminary} 
         \cite{stanca2020impact1} 
         \cite{cainas2021kat}
         \cite{tadesse2022combining} 
         \cite{janvrin2014making}     
         \cite{vidermanova2015visualization} 
         \cite{mendez2015visualizing} 
         \cite{ito2018text} 
         \cite{jonker2019industry}
         \cite{prokofieva2020visualization}
         \cite{ito2020ginn} 
         \cite{choi2023enhancing}
         \cite{stanca2020impact2} 
         \cite{schroder2022pension} 
         \cite{tomasi2023role}
         \cite{hirsch2015visualisation} 
         \cite{lusardi2017visual} 
         \cite{chan2016finavistory}
          \\
        Making investment decisions & 
         \cite{savikhin2008applied} 
         \cite{savikhin2011experimental}
         \cite{wang2024visualization} \\ \hline
        \textbf{Learning Objective Level} & \multicolumn{1}{c}{\textbf{Paper}}
        \\ \hline
        Financial skills of daily living & 
         \cite{kothakota2020use} 
         \cite{csallner2003fundexplorer} 
         \cite{ervin2004visualizing} 
         \cite{rudolph2009finvis} 
         \cite{savikhin2011experimental} 
         \cite{vidermanova2015visualization} 
         \cite{ito2018text} 
         \cite{lugmayr2019financial} 
         \cite{ito2020ginn} 
         \cite{choi2023enhancing}
         \cite{wang2024visualization}
         \cite{schroder2022pension}
         \cite{lusardi2017visual} 
         \cite{chan2016finavistory} 
        \\
        Basic professional skills &  
         \cite{helweg2007business} 
         \cite{savikhin2008applied} 
         \cite{tanlamai2011learning1} 
         \cite{tanlamai2011learning2} 
         \cite{raungpaka2013preliminary}
         \cite{stanca2020impact1} 
         \cite{cainas2021kat}
         \cite{tadesse2022combining}
         \cite{janvrin2014making} 
         \cite{zhang2014incorporating} 
         \cite{mendez2015visualizing}
          \cite{kokina2017role} 
         \cite{staveley2018integrating}
         \cite{schonfeldt2020ict} 
         \cite{prokofieva2020visualization} 
         \cite{laplante2021incorporating} 
         \cite{varma2022web} 
         \cite{kidwell2024tableau}
         \cite{stanca2020impact2} 
         \cite{hirsch2015visualisation}
        \\
        Advanced financial skills & 
         \cite{lemieux2014using} 
         \cite{jonker2019industry} 
         \cite{tomasi2023role} \\ \hline
    \end{tabular}
    }
    \label{knowledge_learning_obj_target}
\end{table*}

For learning objective topics, ``interpreting visualization" dominates (26 papers), particularly in investment-related studies with ``learning to understand concepts" as the objective type. Here, ``interpreting visualization" refers to learning how to read and make sense of financial information presented in visual formats such as charts, graphs, or dashboards. These papers \cite{csallner2003fundexplorer, rudolph2009finvis, tanlamai2011learning1, lusardi2017visual, kothakota2020use, stanca2020impact2, schroder2022pension} rarely prioritize ``making investment decisions" as an objective. Meanwhile, 7 out of 11 papers on ``learning to use tools" primarily focus on ``developing business accounting skills." Only three papers \cite{wang2024visualization, savikhin2008applied, savikhin2011experimental} identify ``making investment decisions" as a key objective.

In terms of learning objective levels, most studies target ``basic professional skills," relevant to roles like bank employees, accountants, and insurance agents. Many also address the general public, focusing on ``financial skills of daily living" (e.g., opening bank accounts, budgeting, and retirement savings). Only three papers \cite{lemieux2014using, jonker2019industry, tomasi2023role} cover ``advanced financial skills," all targeting experts like share traders and investment bankers. Additionally, one paper \cite{hirsch2015visualisation} targets experts but emphasizes ``basic professional skills."

\begin{figure*}[htbp]
    \centering
    \includegraphics[width=0.9\textwidth, height=6cm]{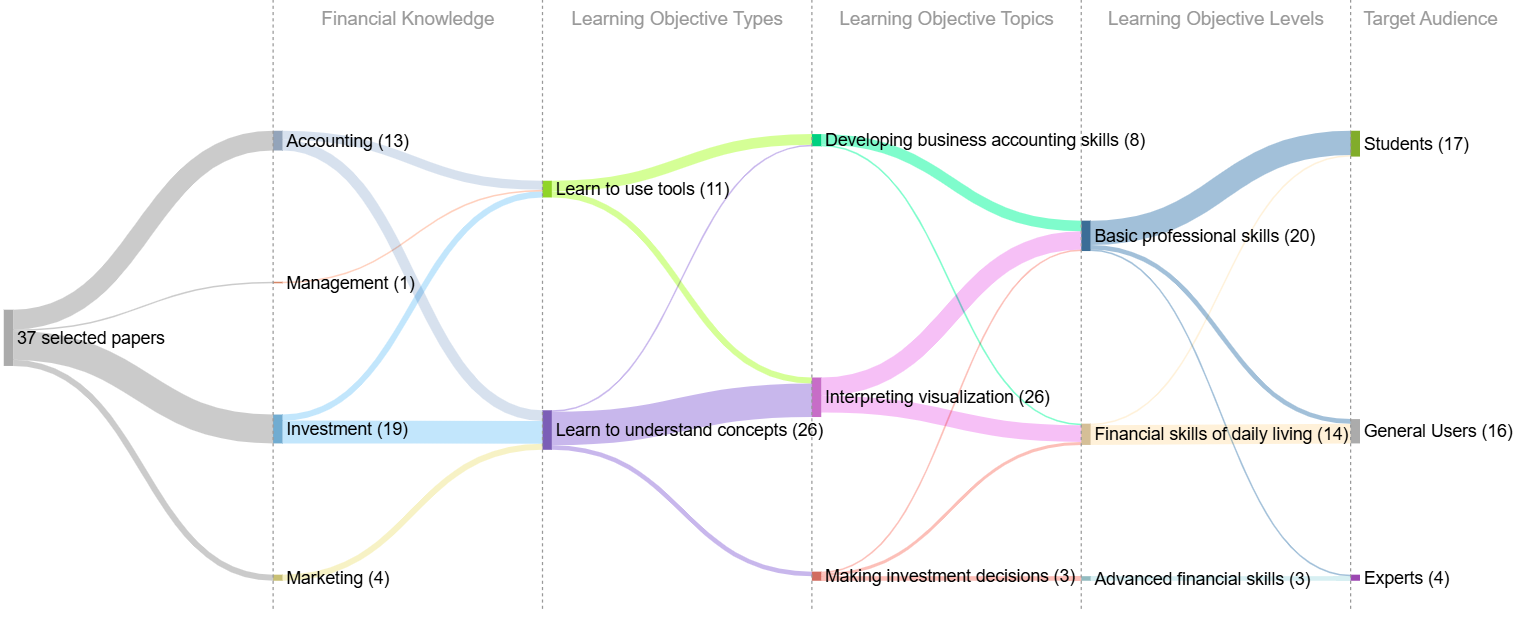}
    \caption{The Sankey diagram highlights the main research pathways, from financial knowledge through learning objectives to target audiences, and exposes gaps where certain skills or audiences are underrepresented.}
    \label{knowledge-type-topic-level-target-audience-sankey}
\end{figure*}

To visualize the predominant patterns among research domains, learning objectives, and target audiences, we created a Sankey diagram, as shown in Figure \ref{knowledge-type-topic-level-target-audience-sankey}. From the diagram, it is clear that the most common research focus follows this pathway: \setulcolor{LightSkyBlue}
\setul{1.5pt}{1.5pt}\ul{Investment} - \setulcolor{MediumPurple}\setul{1.5pt}{1.5pt}\ul{Learning to Understand Concepts} - \setulcolor{Violet}\setul{1.5pt}{1.5pt}\ul{Interpreting Visualization} - \setulcolor{SteelBlue}\setul{1.5pt}{1.5pt}\ul{Basic Professional Skills} - \setulcolor{YellowGreen}\setul{1.5pt}{1.5pt}\ul{Students}. This pattern highlights a strong emphasis on foundational financial knowledge tailored for students, with a particular focus on developing their ability to interpret visualizations in the context of investments. However, it also reveals a potential gap: researchers give limited attention to advanced financial skills and more diverse target audiences, such as industry professionals or the general public. This may be due to the complexity of advanced financial topics, which are harder to communicate visually, or the traditional academic focus on student-centered educational settings.

\subsubsection{Learning barriers and pedagogy}

Some papers referred to ``learning challenges" or provided descriptions related to the potential difficulties the students encountered, but they did not specifically use the term ``learning barriers." Similarly, while the term ``pedagogy" was rarely directly mentioned, the authors discussed teaching methods such as project-based learning, team-based collaborative learning, and scaffolded learning. Based on this information, we identified four types of learning barriers and five pedagogical methods, as shown in Table \ref{learning_barriers_pedagogy}. 

\begin{table*}[htbp]
    \centering
    \renewcommand{\arraystretch}{1.5}
    \caption{Categories of Learning Barriers and Pedagogical Methods}
    \resizebox{\textwidth}{!}{
    \begin{tabular}{>{\centering\arraybackslash}m{5.5cm} >{\raggedright\arraybackslash}m{12.5cm}}
        \hline
        \textbf{Learning Barrier} & \multicolumn{1}{c}{\textbf{Paper}}
        \\ \hline
        Data Complexity and Technical Challenges & 
        \cite{csallner2003fundexplorer} 
        \cite{rudolph2009finvis} 
        \cite{savikhin2011experimental} 
        \cite{tanlamai2011learning1} 
        \cite{janvrin2014making} 
        \cite{zhang2014incorporating}
        \cite{lemieux2014using} 
        \cite{hirsch2015visualisation} 
        \cite{chan2016finavistory} 
        \cite{lugmayr2019financial} 
        \cite{jonker2019industry} 
        \cite{tadesse2022combining} 
        \cite{varma2022web} 
        \cite{schroder2022pension}
        \cite{choi2023enhancing}\\
        Cognitive Load and Resistance to Change & 
        \cite{savikhin2008applied} 
        \cite{rudolph2009finvis} 
        \cite{savikhin2011experimental} 
        \cite{tanlamai2011learning1} 
        \cite{raungpaka2013preliminary} 
        \cite{stanca2020impact1} 
        \cite{stanca2020impact2} 
        \cite{schroder2022pension} 
        \cite{tomasi2023role}
        \cite{choi2023enhancing} \\
        Limited Financial Knowledge and Data Skills & 
        \cite{ervin2004visualizing} 
        \cite{helweg2007business} 
        \cite{tanlamai2011learning2} 
        \cite{zhang2014incorporating} 
        \cite{vidermanova2015visualization} 
        \cite{mendez2015visualizing} 
        \cite{lusardi2017visual} 
        \cite{kokina2017role} 
        \cite{ito2018text} 
        \cite{staveley2018integrating} 
        \cite{lugmayr2019financial} 
        \cite{kothakota2020use} 
        \cite{prokofieva2020visualization} 
        \cite{ito2020ginn} 
        \cite{cainas2021kat}
        \cite{kidwell2024tableau}
        \cite{wang2024visualization}\\
        Lack of Practical Experience & 
        \cite{raungpaka2013preliminary} 
        \cite{hirsch2015visualisation} 
        \cite{schonfeldt2020ict} 
        \cite{stanca2020impact2}
        \cite{laplante2021incorporating} 
        \cite{choi2023enhancing}
        \\
        \hline
        \textbf{Pedagogical Method} & \multicolumn{1}{c}{\textbf{Paper}}
        \\ \hline
        Interactive Learning & 
        \cite{csallner2003fundexplorer} 
        \cite{savikhin2008applied} 
        \cite{rudolph2009finvis} 
        \cite{savikhin2011experimental} 
        \cite{janvrin2014making} 
        \cite{zhang2014incorporating} 
        \cite{lemieux2014using} 
        \cite{mendez2015visualizing} 
        \cite{chan2016finavistory} 
        \cite{lusardi2017visual} 
        \cite{kokina2017role}
        \cite{staveley2018integrating} 
        \cite{lugmayr2019financial}
        \cite{jonker2019industry} 
        \cite{schonfeldt2020ict} 
        \cite{stanca2020impact1} 
        \cite{laplante2021incorporating}
        \cite{varma2022web} 
        \cite{choi2023enhancing}
        \cite{wang2024visualization} \\
        Project-Based Learning & 
        \cite{tanlamai2011learning1} 
        \cite{janvrin2014making} 
        \cite{vidermanova2015visualization} 
        \cite{mendez2015visualizing} 
        \cite{kokina2017role} 
        \cite{prokofieva2020visualization} 
        \cite{cainas2021kat}
        \cite{laplante2021incorporating} 
        \cite{tadesse2022combining}
        \cite{varma2022web} 
        \cite{kidwell2024tableau}\\
        Collaborative Learning & 
        \cite{schonfeldt2020ict} 
        \cite{stanca2020impact2} \\
        Scaffolded Learning & 
        \cite{staveley2018integrating}
        \cite{schonfeldt2020ict} 
        \cite{prokofieva2020visualization}
        \\
        Traditional Learning & 
        \cite{ervin2004visualizing} 
        \cite{helweg2007business} 
        \cite{tanlamai2011learning1} 
        \cite{tanlamai2011learning2} 
        \cite{raungpaka2013preliminary} 
        \cite{hirsch2015visualisation} 
        \cite{ito2018text} 
        \cite{kothakota2020use} 
        \cite{ito2020ginn} 
        \cite{schroder2022pension} 
        \cite{tomasi2023role} \\
        \hline
    \end{tabular}
    }
    \label{learning_barriers_pedagogy}
\end{table*}

\begin{figure}[htbp]
    \centering
    \includegraphics[width=\columnwidth]{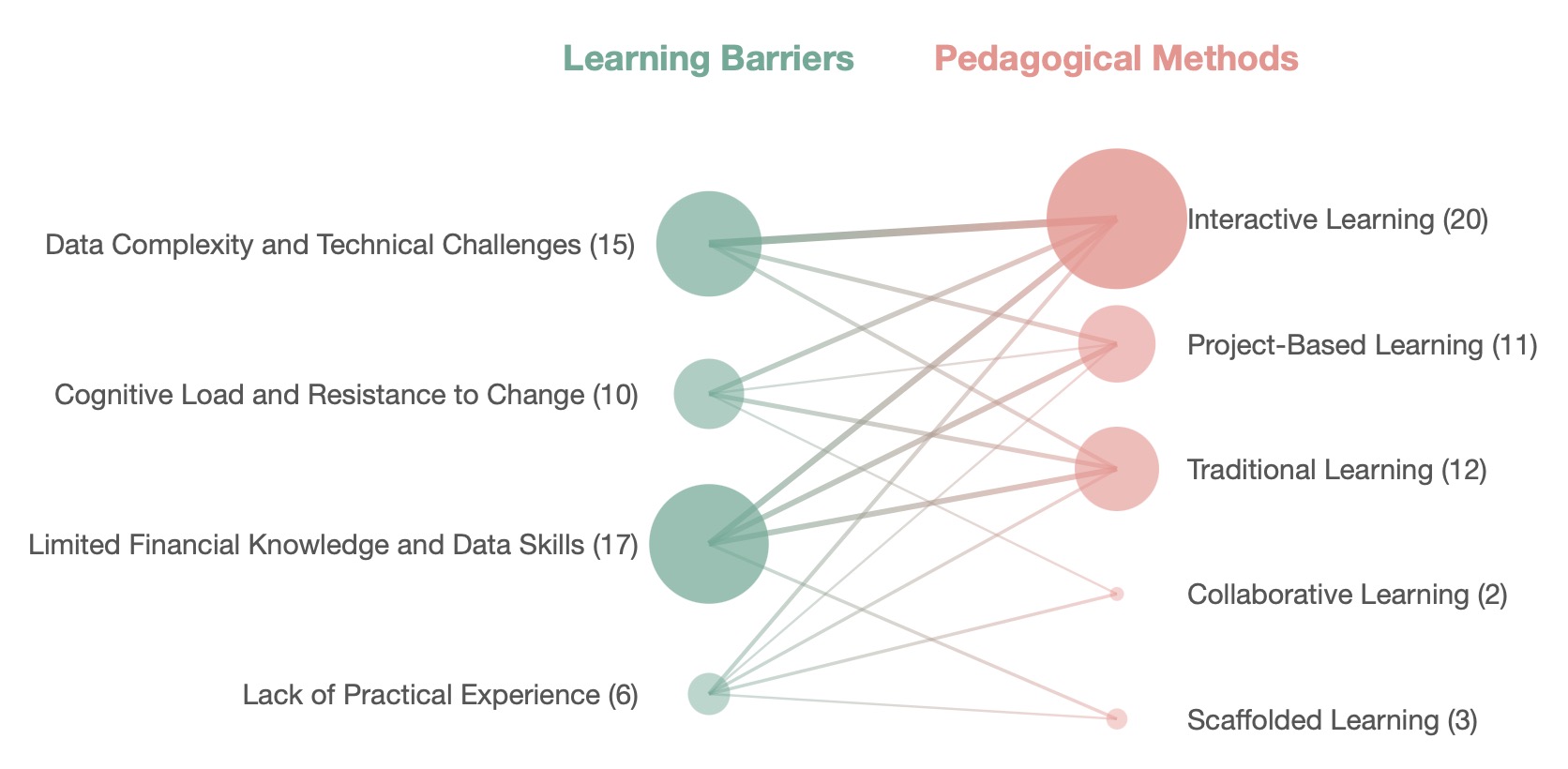}
    \caption{The bipartite graph illustrates which teaching methods are most frequently used to tackle each learning barrier, with larger nodes showing more papers per category and thicker arcs indicating stronger method–barrier links.}
    \label{learning-barrier-pedagogy-bipartite-graph}
\end{figure}

We identified patterns linking learning barriers to pedagogical methods, showing how specific approaches address particular challenges, as visualized in the Arc Diagram presented in Figure \ref{learning-barrier-pedagogy-bipartite-graph}. Each node represents a specific concept, colored in Pine Green for learning barriers and Peach for pedagogical methods. The radius and transparency of each node reflect the number of papers in that category—larger nodes with lower transparency indicate more papers. This quantity is also shown in parentheses within the label of each node for clarity. Arcs represent how specific pedagogical methods are used to address particular learning barriers. Their thickness indicates the strength or frequency of each connection, while a color gradient—shifting from Pine Green at the source (learning barriers) to Peach at the target (pedagogical methods)—visually conveys the direction of the relationship.

The results from Figure \ref{learning-barrier-pedagogy-bipartite-graph} and Table \ref{learning_barriers_pedagogy} show that 10 papers address multiple learning barriers and 9 explored multiple pedagogical methods. For example, Choi and Kim \cite{choi2023enhancing} addressed three different barriers, while Tanlamai and Soongswang \cite{tanlamai2011learning1} combined project-based and traditional methods. Notably, certain pedagogical approaches show clear preferences in specific contexts. Data Complexity and Technical Challenges is linked to Interactive Learning in 11 papers, such as \cite{csallner2003fundexplorer, rudolph2009finvis, savikhin2011experimental, varma2022web}, suggesting that student-centered strategies help learners navigate complex content. Cognitive Load and Resistance to Change is also frequently addressed with Interactive Learning (5 papers), such as \cite{savikhin2008applied, rudolph2009finvis, stanca2020impact1}, indicating that flexible instruction supports adaptation. Among the 17 papers addressing Limited Financial Knowledge and Data Skills, 7 use Interactive Learning and 6 adopt Project-Based Learning—the two most common strategies for this barrier—as seen in \cite{mendez2015visualizing, kokina2017role}. Only one paper \cite{schonfeldt2020ict} applied Scaffolded Learning to address Lack of Practical Experience, suggesting its potential to bridge experience gaps, though further validation is needed.
\subsubsection{Feedback mechanism in the Learning Process}

\begin{figure*}[htbp]
    \centering
    \makebox[\textwidth]{
        \includegraphics[width=\textwidth]{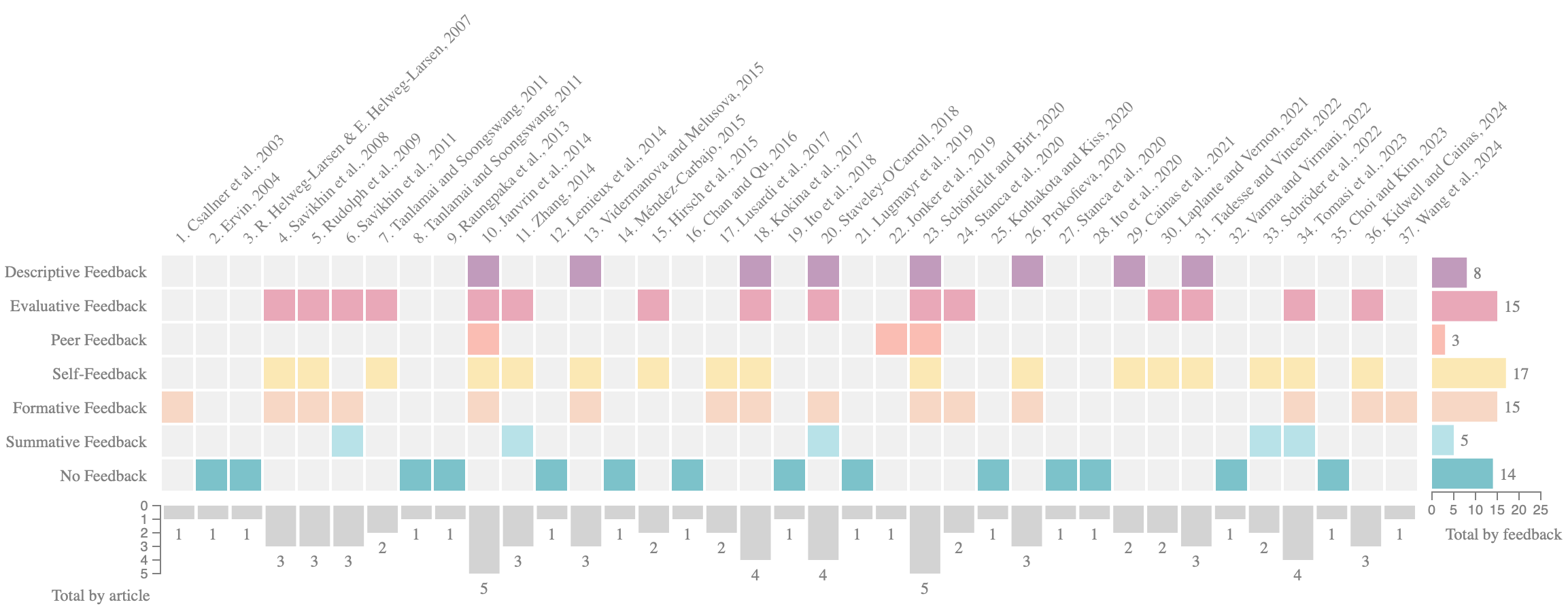}
    }
    \caption{Color coding and histograms of feedback type distribution. Color coding distinguishes seven feedback types. The right histogram shows how many papers include each type, and the bottom histogram shows how many feedback types each paper addresses.}
    \label{feedback-heatmap-new-order-bottom-right-histogram.png}
\end{figure*}

The feedback learners receive during the learning process is essential for guiding their progress, reinforcing understanding, and addressing misconceptions effectively. We identified seven feedback types and classified the papers accordingly, as shown in Figure \ref{feedback-heatmap-new-order-bottom-right-histogram.png}, where horizontal and vertical histograms show feedback types per paper and paper distribution across types.

Descriptive Feedback provides written, specific details about performance; Evaluative Feedback involves performance judgments such as grades; Peer Feedback is given by classmates or other learners; Self-Feedback encourages learners to reflect on their own progress; Formative Feedback is ongoing, guiding improvements during the learning process; Summative Feedback summarizes performance at the end of a task or unit; and No Feedback indicates that the papers do not address feedback.

The vertical histogram at the bottom shows that 20 papers addressed multiple feedback types (i.e., counts above one). For example, Savikhin et al. discussed evaluative, self-, and formative feedback \cite{savikhin2008applied}, emphasizing the value of diverse feedback in learning. Two papers covered nearly all types \cite{janvrin2014making, schonfeldt2020ict}, reflecting comprehensive and varied feedback strategies.

The horizontal histogram on the right shows that 14 papers mentioned no feedback types, while self-feedback appeared most frequently (17 papers). Formative and evaluative feedback were also common, each cited in 15 papers, highlighting the importance of ongoing support and assessment. In contrast, peer and summative feedback were less common, appearing in only 3 and 5 papers, possibly because these studies focused more on continuous support than on peer interaction or post-learning evaluation.

\subsection{What visualization tools and technologies are used in financial education and why?}

Visualization tools and technologies in financial education play a key role in enhancing learning outcomes and improving financial literacy. In this section, we address five aspects. First, we examine the software used, including the reasons for its adoption and its strengths and limitations. Second, we review the datasets used to generate visualizations. Third, we discuss the types of visualizations and the rationale for their selection. Fourth, we evaluate interaction design and its benefits. Finally, we analyze the task domain and provide examples of practical applications.

\subsubsection{Visualization Tools for Financial Education}

Table \ref{visualization_tool} presents the visualization tools identified in the 37 selected papers, categorized into commercial tools and self-developed tools.

\begin{table*}[htbp]
    \centering
    \renewcommand{\arraystretch}{1.2}
    \caption{Summary of Visualization Tools}
    \resizebox{\textwidth}{!}{
    \begin{tabular}{>{\centering\arraybackslash}m{1.6cm}>{\centering\arraybackslash}m{1cm}>{\raggedright\arraybackslash}m{7cm}>{\centering\arraybackslash}m{2.5cm}>{\centering\arraybackslash}m{2.6cm}>{\centering\arraybackslash}m{2.6cm}>{\centering\arraybackslash}m{2.8cm}>{\centering\arraybackslash}m{3cm}>{\centering\arraybackslash}m{2.8cm}}
        \hline
        \multirow{2}[3]{1.6cm}{\centering\arraybackslash\textbf{Tool Type}} & 
        \multirow{2}[3]{1cm}{\centering\arraybackslash\textbf{Paper}} & 
        \multirow{2}[3]{7cm}{\centering\arraybackslash\textbf{Tool Name}} & 
        \multicolumn{6}{c}{\textbf{Reason for Choosing}} \\ \cline{4-9}
        & & & \textbf{For Industry Popularity} & 
        \textbf{For Tools Accessibility} & 
        \textbf{For Student Familiarity} & 
        \textbf{For Learning and Career Development} & 
        \textbf{For Platform and System Compatibility} & 
        \textbf{For the Functionality} \\ 
        \hline
        \multirow{12}[20]{1.6cm}{\centering Commercial Tools}
        & \cite{cainas2021kat} & Excel, Power BI, Tableau & \texttimes & \texttimes & \texttimes & \texttimes & \texttimes & \texttimes\\
        & \cite{tadesse2022combining} & Tableau Desktop, Power BI, QlikView, Excel with add-ons Power Pivot & \texttimes & & & & & \\
        & \cite{janvrin2014making} & Interactive Data Visualization (IDV) Software: Tableau, AnalystX Office, Centrifuge, Spotfire, Visual Mining, Excel Dashboards & & & & & &  \texttimes \\ 
        & \cite{zhang2014incorporating} & Excel tools: Data Table and Chart, Scenario Manager, Goal Seek and Solver. & & & & & & \texttimes  \\
        & \cite{mendez2015visualizing} & FRED (Federal Reserve Economic Data), an online database offers tools for presenting data. & & \texttimes & & & &  \texttimes \\ 
        & \cite{kokina2017role} & Excel, Tableau & & & \texttimes & & &   \\ 
        & \cite{staveley2018integrating} & FRED (Federal Reserve Economic Data), an online database offers tools for presenting data. & & & \texttimes & \texttimes & &  \\ 
        & \cite{jonker2019industry} & Excel, Xero, Tableau & & & & \texttimes & & \texttimes\\ 
        & \cite{prokofieva2020visualization} & Tableau & & & & \texttimes & & \texttimes\\ 
        & \cite{laplante2021incorporating} & Excel, Tableau & \texttimes & & & & & \texttimes \\
        & \cite{varma2022web} & Shiny package developed by Rstudio. & & & & & \texttimes &  \texttimes \\ 
        & \cite{kidwell2024tableau} & Tableau & & & & \texttimes & & \texttimes  \\ 
        \cline{2-4}
        \hline
        SUM &  &  & 3 & 2 & 3 & 5 & 2 & 9  \\ 
        \hline

        &  &  & \multicolumn{3}{c}{\textbf{Addressing Limitations of Existing Methods}} & \multicolumn{3}{c}{\textbf{For Supporting Research Purposes}} \\ 
        \cline{4-9}
        
        \multirow{15}[5]{1.6cm}{\centering Self-developed Tools}
        & \cite{csallner2003fundexplorer} & Visual analytics tool, FundExplorer & \multicolumn{3}{c}{\texttimes} & \multicolumn{3}{c}{}\\ 
        & \cite{helweg2007business} & Visualization tool & \multicolumn{3}{c}{\texttimes} & \multicolumn{3}{c}{}\\ 
        & \cite{savikhin2008applied} & Visual analytics tool & \multicolumn{3}{c}{} & \multicolumn{3}{c}{\texttimes} \\ 
        & \cite{rudolph2009finvis} & Visual analytics tool, FinVis & \multicolumn{3}{c}{\texttimes} & \multicolumn{3}{c}{} \\ 
        & \cite{lemieux2014using} & Visual analytics tool & \multicolumn{3}{c}{\texttimes} & \multicolumn{3}{c}{}\\ 
        & \cite{raungpaka2013preliminary} & Visualization with metaphor, Lotus Viz & \multicolumn{3}{c}{} & \multicolumn{3}{c}{\texttimes} \\ 
        & \cite{savikhin2011experimental} & Visual analytics tool, PortfolioCompare & \multicolumn{3}{c}{\texttimes} & \multicolumn{3}{c}{}\\ 
        & \cite{ito2018text} & Gradient Interpretable Neural Network (GINN) Architecture & \multicolumn{3}{c}{} & \multicolumn{3}{c}{\texttimes} \\ 
        & \cite{lugmayr2019financial} & 3D tool, ElectrAus & \multicolumn{3}{c}{} & \multicolumn{3}{c}{\texttimes} \\ 
        & \cite{jonker2019industry} & Visual analytics approach & \multicolumn{3}{c}{\texttimes} & \multicolumn{3}{c}{}\\ 
        & \cite{ito2020ginn} & Gradient Interpretable Neural Network (GINN) Architecture & \multicolumn{3}{c}{} & \multicolumn{3}{c}{\texttimes} \\ 
        & \cite{wang2024visualization} & Visual analytics tool & \multicolumn{3}{c}{\texttimes} & \multicolumn{3}{c}{}\\
        & \cite{schroder2022pension} & Data-driven storytelling application & \multicolumn{3}{c}{} & \multicolumn{3}{c}{\texttimes} \\ 
        & \cite{lusardi2017visual} & Visual analytics tool, FinVis & \multicolumn{3}{c}{\texttimes} & \multicolumn{3}{c}{}\\ 
        & \cite{chan2016finavistory} & Visual narrative system, FinaVistory & \multicolumn{3}{c}{\texttimes} & \multicolumn{3}{c}{\texttimes} \\ 
        \cline{2-4} 
        \hline
        SUM &  & & \multicolumn{3}{c}{9} & \multicolumn{3}{c}{7} \\  
        \hline
        
        \multirow{3}{1.6cm}{\centering Unspecified}
        &   \multirow{3}{8.5cm}{
        \cite{kothakota2020use} \cite{ervin2004visualizing}
        \cite{tanlamai2011learning1} 
        \cite{tanlamai2011learning2} 
        \cite{stanca2020impact1}
        \cite{vidermanova2015visualization}
        \cite{choi2023enhancing}
        \cite{stanca2020impact2} 
        \cite{tomasi2023role}
        \cite{hirsch2015visualisation}
        } & & \multicolumn{6}{l}{The visualization tools or methods used are not clearly identified or described in the study.} \\
        &&& \multicolumn{6}{l}{Two papers \cite{tanlamai2011learning2, vidermanova2015visualization} mention tools as ``Half self-developed, half borrowed from the internet," with no} \\ 
        &&& \multicolumn{6}{l}{further implementation details provided.} \\ 
        
        \hline
    \end{tabular}
    }
    \label{visualization_tool}
\end{table*}

Commercial tools were used in twelve papers, with popular options including Tableau, Excel, Power BI, and others \cite{kokina2017role, schonfeldt2020ict, cainas2021kat, tadesse2022combining}. Additional tools included the economic database FRED \cite{mendez2015visualizing, staveley2018integrating}, the R-based package Shiny \cite{varma2022web}, and the cloud accounting platform Xero \cite{schonfeldt2020ict}. The primary reasons for selecting commercial tools were their functionality (noted in nine papers) and their contribution to students’ learning and career readiness (highlighted in five papers). Notably, Cainas et al. \cite{cainas2021kat} offered a comprehensive evaluation, considering not only functionality and learning outcomes but also software accessibility, industry popularity, student familiarity, and compatibility.

Self-developed tools appeared in fifteen papers, often addressing limitations of existing tools or targeting specific research goals. Some, such as \cite{rudolph2009finvis, lusardi2017visual}, compared custom tools with other visual representations, while others created tools for specific educational purposes. For example, Savikhin et al. \cite{savikhin2008applied} illustrated the Winner’s and Loser’s Curse, Lugmayr et al. \cite{lugmayr2019financial} built a 3D tool for immersive storytelling, and Raungpaka et al. \cite{raungpaka2013preliminary} used natural metaphors like lotus plants to visualize accounting information.
\subsubsection{Dataset for Visualization Generation}

Due to the lack of detailed data descriptions or visualization figures in many papers, it is difficult to classify the datasets as time-series, geospatial, network, or textual. Instead, we grouped them into three categories based on financial characteristics and usage: Share Market Data (17 papers), Financial Accounting Data (15 papers), and Management Accounting Data (5 papers), as shown in Table \ref{visualization_type}.

\begin{table*}[htbp]
    \centering
    \renewcommand{\arraystretch}{1.2} 
    \caption{Summary of Datasets and Visualization Types}
    \resizebox{\textwidth}{!}{
    \begin{tabular}{>{\raggedright\arraybackslash}m{1cm}>{\centering\arraybackslash}m{6cm}>{\centering\arraybackslash}m{8cm}>{\centering\arraybackslash}m{4cm}>{\centering\arraybackslash}m{4cm}>
    {\centering\arraybackslash}m{6.5cm}}
        \hline
        \multirow{2}[1]{1cm}{\centering\arraybackslash\textbf{Paper}} & 
        \multirow{2}[1]{6cm}{\centering\arraybackslash\textbf{Dataset}} & 
        \multicolumn{3}{c}{\textbf{Visualization Type}} 
         &
        \multirow{2}[1]{6.5cm}{\centering\arraybackslash\textbf{Reason for choosing}}
        \\ 
        \cline{3-5}
        & & 
        \textbf{2D Standard/Dashborad} & 
        \textbf{3D} & 
        \textbf{2D Self-Designed}\\ 
        \hline
        \cite{csallner2003fundexplorer} & Share Market Data & Treemap & & & Exploratory  \\ 
        \cite{ervin2004visualizing} & Share Market Data & Pie Chart, Scatter Plot & & & Exploratory, Descriptive, Predictive  \\ 
        \cite{rudolph2009finvis} & Share Market Data & ThemeRiver Visualization & & & Exploratory  \\ 
        \cite{savikhin2011experimental} & Share Market Data & Bar Chart, Scatter Plot & & & Explanatory \\
        \cite{lemieux2014using} & Share Market Data & Treemap, Bar Chart, Matrix Visualization & & & Exploratory  \\ 
        \cite{mendez2015visualizing} & Share Market Data & Line Chart & & & Descriptive  \\ 
        \cite{chan2016finavistory} & Share Market Data & Network Graph, Bubble Chart, Stacked Bar Chart, Word Cloud  & & & Exploratory   \\
        \cite{lusardi2017visual} & Share Market Data & ThemeRiver Visualization & & Risk cone & Exploratory  \\ 
        \cite{ito2018text} & Share Market Data & & & Text-based & Descriptive  \\ 
        \cite{staveley2018integrating} & Share Market Data & Line Chart & & & Explanatory  \\
        \cite{lugmayr2019financial} & Share Market Data & & ElectrAus & & Predictive\\
        \cite{jonker2019industry} & Share Market Data & Bubble Chart, Choropleth Map, Line Chart & 3D Bar Chart & & Explanatory, Predictive  \\ 
        \cite{ito2020ginn} & Share Market Data & & & Text-based & Descriptive \\ 
        \cite{varma2022web} & Share Market Data & Line Chart & & & Exploratory   \\
        \cite{tomasi2023role} & Share Market Data & Line Chart & & & Explanatory  \\
        \cite{choi2023enhancing} & Share Market Data & Network Graph & & & Exploratory \\ 
        \cite{wang2024visualization} & Share Market Data & Line Chart, Bar Chart, Scatter Plot, Treemap, Heatmap & & & Exploratory, Descriptive  \\ 
        \cite{helweg2007business} & Financial Accounting Data & & The Company Board & & Exploratory \\ 
        \cite{tanlamai2011learning1} & Financial Accounting Data & Bar Chart, Stacked Bar Chart & & & Explanatory, Predictive \\ 
        \cite{tanlamai2011learning2} & Financial Accounting Data & Bar Chart, Stacked Bar Chart & & & Predictive \\
        \cite{raungpaka2013preliminary} & Financial Accounting Data & & & Lotus Flower Metaphor & Explanatory\\ 
        \cite{zhang2014incorporating} & Financial Accounting Data & Line Chart & & & Explanatory, Predictive  \\
        \cite{vidermanova2015visualization} & Financial Accounting Data & Line Chart & & & Explanatory  \\ 
        \cite{schonfeldt2020ict} & Financial Accounting Data & Dashboard & & & Exploratory, Predictive  \\ 
        \cite{stanca2020impact1} & Financial Accounting Data & Dashboard & & & Explanatory  \\
        \cite{kothakota2020use} & Financial Accounting Data & Bar Chart, Stacked Bar Chart & & & Explanatory \\ 
        \cite{prokofieva2020visualization} & Financial Accounting Data & Dashboard & & & Explanatory  \\ 
        \cite{cainas2021kat} & Financial Accounting Data & Choropleth Map, Pivot Chart & &   & Explanatory, Descriptive   \\
        \cite{laplante2021incorporating} & Financial Accounting Data & Unspecified Graphs and Figures by Tableau & & &  Exploratory, Explanatory \\ 
        \cite{tadesse2022combining} & Financial Accounting Data & Unspecified Visualization by Tableau& & & Explanatory \\ 
        \cite{schroder2022pension} & Financial Accounting Data & & & Dots Animation &  Explanatory \\ 
        \cite{kidwell2024tableau} & Financial Accounting Data & Bar Chart, Scatter Plot, Treemap, Stacked Bar & & & Exploratory   \\
        \cite{savikhin2008applied} & Management Accounting Data & Line Chart & & & Explanatory, Predictive\\ 
        \cite{janvrin2014making} & Management Accounting Data & Line Chart, Bar Chart, Pie Chart & & & Exploratory   \\
        \cite{hirsch2015visualisation} & Management Accounting Data & Bar Chart & & & Explanatory   \\
        \cite{kokina2017role} & Management Accounting Data & Line Chart, Scatter Plot, Heatmap & & & Exploratory \\
        \cite{stanca2020impact2} & Management Accounting Data & Dashboard & & & Explanatory   \\
        \hline
        SUM &  & 31 & 3 & 5 &   \\ 
        \hline
    \end{tabular}
    }
    \label{visualization_type}
\end{table*}


Share Market Data includes stock prices, trading volume, market capitalization, and dividend yield, revealing trends and investment potential. Financial Accounting Data covers financial statements, standards, analysis, and auditing, supporting transparency and evaluation. Management Accounting Data aids internal decisions through cost control, budgeting, forecasting, and performance tracking to improve planning and resource use.
\subsubsection{Visualization Types for Representation}

We categorized visualization types into three groups: (1) 2D standard visualizations and dashboards, such as bar, line, and pie charts; (2) 2D self-designed visualizations, which deviate from conventional chart types and are custom-built by the authors; and (3) 3D visualizations, including immersive or spatial formats. As shown in Table \ref{main_table_5_year_trend}, publications using both 2D standard and 2D self-designed visualizations have increased, while 3D usage has remained limited and inconsistent over time.



As shown in Table \ref{visualization_type}, 31 papers adopted 2D visualizations, primarily using standard forms such as line, bar, and scatter plots. Four of these papers incorporated dashboards built with tools like Tableau \cite{schonfeldt2020ict, stanca2020impact1, prokofieva2020visualization, stanca2020impact2}, while two papers mentioned Tableau without specifying the chart types used \cite{tadesse2022combining, laplante2021incorporating}.

Only three papers employed 3D visualizations, each with different motivations: to support non-expert understanding through a dynamic landscape \cite{helweg2007business}, to improve model transparency with 3D bar charts \cite{jonker2019industry}, and to explore immersive storytelling in the energy sector \cite{lugmayr2019financial}. 

Five papers featured 2D self-designed visualizations, such as the ThemeRiver and a “risk cone” for illustrating risk \cite{rudolph2009finvis, lusardi2017visual}, text-based visualizations for financial sentiment \cite{ito2018text, ito2020ginn}, a lotus-inspired design for business ecosystems \cite{raungpaka2013preliminary}, and animated dots to show pension fund growth \cite{schroder2022pension}.

We analyzed the rationale behind selecting specific visualization types for conveying financial knowledge, categorizing them into four types: exploratory, explanatory, descriptive, and predictive (Table \ref{visualization_type}).


Explanatory visualizations prioritize clarity by breaking down complex processes, such as mortgage repayment \cite{vidermanova2015visualization}, into understandable narratives. Exploratory visualizations help users identify patterns and trends through open-ended data interaction. Both aim to educate and support independent analysis.
Descriptive visualizations are used to summarize and present existing data in an accessible format without emphasizing causality or deeper insights. Predictive visualizations, on the other hand, are designed to forecast future outcomes or model hypothetical scenarios using historical or real-time data. Though less common, descriptive and predictive visualizations contribute by offering straightforward summaries or foresight-oriented insights to support financial decision-making.
\subsubsection{Interaction Modes for Exploration}


Although many papers did not explicitly specify interaction modes, we inferred them from keywords such as “click,” “select,” “highlight,” “filter,” “sort,” “manage,” and “explore,” and categorized them into eight types, shown as Table \ref{interaction mode}. The most common interactions were data-centric, led by Data Exploration and Navigation (11 papers), which helps uncover patterns and deepen understanding. This was followed by Data Operation and Management (10 papers) and Data Selection and Highlighting (9 papers), which support analytical efficiency by focusing attention on key attributes. In contrast, other types like Filtering and Sorting (6 papers) and Multi-View Coordination (3 papers) were less frequent.

\begin{table*}[htbp]
    \centering
    \renewcommand{\arraystretch}{1.2}
    \caption{Summary of Interaction Modes}
    \resizebox{1\textwidth}{!}{
    \begin{tabular}{>{\raggedright\arraybackslash}m{1cm}>{\centering\arraybackslash}m{3cm}>{\centering\arraybackslash}m{3cm}>
    {\centering\arraybackslash}m{3cm}>{\centering\arraybackslash}m{3cm}>{\centering\arraybackslash}m{3cm}>
    {\centering\arraybackslash}m{3cm}>
    {\centering\arraybackslash}m{3cm}>{\centering\arraybackslash}m{3cm}}
        \hline
        
        \multirow{2}[2]{1cm}{\centering\arraybackslash\textbf{Paper}} & 
        \multicolumn{8}{c}{\textbf{Interactive Type}} \\ \cline{2-9}
        & \textbf{Data Exploration and Navigation} & 
        \textbf{Data Selection and Highlighting} & 
        \textbf{Data Operation and Management} & 
        \textbf{Filtering and Sorting} & 
        \textbf{Multi-View Coordination} & 
        \textbf{Unspecified} &
        \textbf{No interaction} &
        \textbf{Unknown} 
        \\ 
        \hline
        
        \cite{kothakota2020use} & & & & & & & \texttimes & \\ 
        \cite{csallner2003fundexplorer} & \texttimes & \texttimes & \texttimes & \texttimes & & && \\ 
        \cite{ervin2004visualizing} & & & & & & & \texttimes & \\ 
        \cite{helweg2007business} & & & & & & & \texttimes & \\ 
        \cite{savikhin2008applied} & \texttimes & \texttimes & \texttimes & & & && \\
        \cite{rudolph2009finvis} & \texttimes & \texttimes & \texttimes & \texttimes & \texttimes & && \\ 
        \cite{lemieux2014using} & & & \texttimes & \texttimes & & && \\ 
        \cite{tanlamai2011learning1} & & & & & & & \texttimes & \\ 
        \cite{tanlamai2011learning2} & & & & & & & \texttimes & \\ 
        \cite{raungpaka2013preliminary} & & & & & & & \texttimes & \\
        \cite{stanca2020impact1} & & & & & & && \texttimes \\
        \cite{cainas2021kat} & & & & & & \texttimes && \\
        \cite{tadesse2022combining} & & & & & & \texttimes && \\ 
        \cite{savikhin2011experimental} & \texttimes & \texttimes & \texttimes & & & && \\ 
        \cite{janvrin2014making} & & & & & & \texttimes && \\
        \cite{zhang2014incorporating} & & & & & & \texttimes && \\
        \cite{vidermanova2015visualization} & \texttimes & & & & & & & \\ 
        \cite{mendez2015visualizing} & \texttimes & & \texttimes & & & && \\ 
        \cite{kokina2017role} & & & & & & \texttimes && \\ 
        \cite{ito2018text} & & & & & & && \texttimes \\
        \cite{staveley2018integrating} & & & & & & \texttimes && \\ 
        \cite{lugmayr2019financial} & \texttimes & & & & & && \\ 
        \cite{jonker2019industry} & \texttimes & & \texttimes & & & && \\ 
        \cite{schonfeldt2020ict} & \texttimes & \texttimes & & \texttimes & \texttimes &  &&\\ 
        \cite{prokofieva2020visualization} & & & & & & \texttimes && \\ 
        \cite{ito2020ginn} & & & & & & && \texttimes \\ 
        \cite{laplante2021incorporating} & & & & & & \texttimes && \\  
        \cite{varma2022web} & & \texttimes & \texttimes & & & && \\ 
        \cite{choi2023enhancing} & & & & & & \texttimes && \\
        \cite{kidwell2024tableau} & \texttimes & \texttimes & & \texttimes & & && \\
        \cite{wang2024visualization} & & & \texttimes & & & && \\
        \cite{stanca2020impact2} & & & & & & \texttimes && \\
        \cite{schroder2022pension} & & & & & & & \texttimes & \\ 
        \cite{tomasi2023role} & & & & & & & \texttimes & \\
        \cite{hirsch2015visualisation} & & & & & & & \texttimes & \\ 
         
        \cite{lusardi2017visual} & & \texttimes & \texttimes & & & && \\
        \cite{chan2016finavistory} & \texttimes & \texttimes & & \texttimes & \texttimes & && \\
        \hline
        SUM & 11 & 9 & 10 & 6 & 3 & 10 & 9 & 3\\ 
        \hline
    \end{tabular}
    }
    \label{interaction mode}
\end{table*}

Some papers span multiple interaction categories, illustrating the integration of diverse modes to support flexible learning and analysis. For example, Rudolph et al. \cite{rudolph2009finvis} used five interaction modes, while three papers\cite{csallner2003fundexplorer, chan2016finavistory, schonfeldt2020ict} each involved four. Among these, Data Exploration and Navigation is the most common. Exploration, selection, and filtering often appear together, enhancing users’ ability to analyze data comprehensively. 

Ten papers fell into the Unspecified category: they mentioned interactive tools but did not specify the interaction types, making it unclear how users were expected to interact with the visualizations. For example, four used Tableau \cite{prokofieva2020visualization, cainas2021kat, laplante2021incorporating, tadesse2022combining}, and one used FRED \cite{staveley2018integrating}.

Nine papers were categorized as No Interaction. These relied entirely on static content, including static images \cite{ervin2004visualizing, tanlamai2011learning1, tanlamai2011learning2} and non-interactive formats such as video or animation \cite{raungpaka2013preliminary, schroder2022pension}.

In three papers, the presence of interaction was Unknown due to insufficient description. These include one dashboard-based study \cite{stanca2020impact1} and two using custom visualizations \cite{ito2018text, ito2020ginn}.



Overly complex designs with too many interaction modes can overwhelm users, reducing usability. Conversely, static or low-interaction visualizations may hinder engagement and limit personalized insights. Therefore, balancing interactivity and simplicity is crucial in visualization design.

\subsubsection{Task Domain Analysis and Representative Examples}

Task domains in financial education reflect specific contexts or goals. Our analysis identified four domains: building basic financial literacy (16 papers), optimizing investment strategies (4), decoding financial contexts (10), and interpreting complex financial systems (7).

\begin{figure}[htbp]  
    \centering
    \includegraphics[width=0.8\columnwidth]{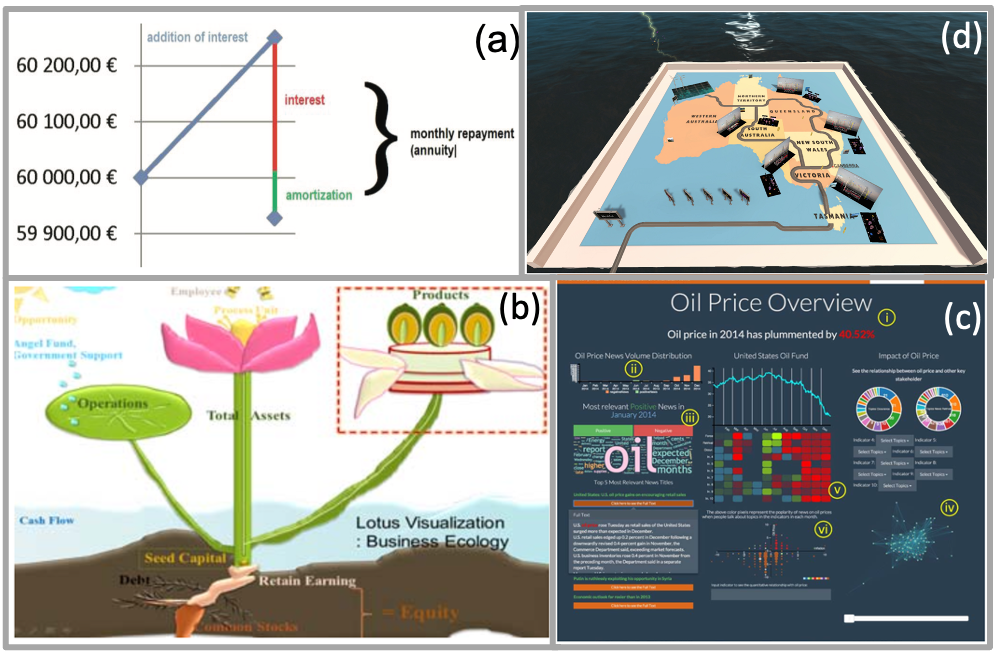} 
    \caption{Representative Examples. (a) The line-based graph breaks down monthly repayments into interest and amortization, offering an intuitive understanding of financial concepts \cite{vidermanova2015visualization}. (b) A lotus flower metaphor maps financial statement elements, enhancing engagement and comprehension \cite{raungpaka2013preliminary}. (c) FinaVistory connects financial news, socio-economic themes, and price dynamics through narrative visualizations \cite{chan2016finavistory}. (d) Immersive 3D visualizations in ``ElectrAus" depict electricity demand, generation, and prices, making complex financial and energy data accessible \cite{lugmayr2019financial}. (Reproduced from these works, © 2013–2019, with kind permission of the authors.)
    }
    \label{example1}
\end{figure}

\paragraph{Building Basic Financial Literacy}
Building basic financial literacy helps individuals grasp essential financial concepts and manage everyday financial tasks. We identified two subcategories: understanding basic concepts, such as mortgages, loans, and repayments, and understanding financial statements and reports, including balance sheets and income statements.

\begin{figure}[htbp]  
    \centering
    \includegraphics[width=0.8\columnwidth]{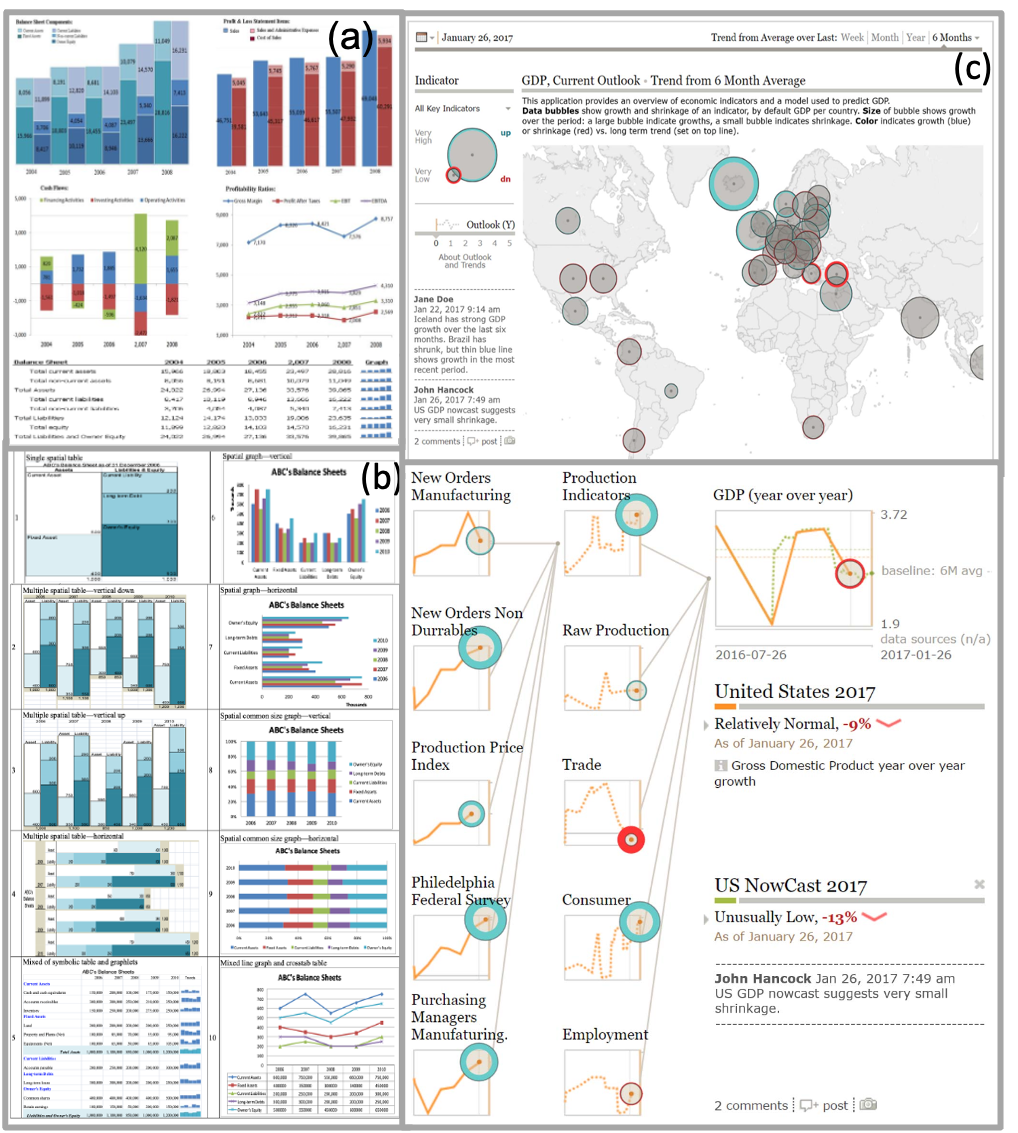} 
    \caption{Representative Examples. (a) “Graphs with numbers” and “tables with graphs” show financial status and risk \cite{tanlamai2011learning1}. (b) Mixed formats—numbers with graphs, spatial tables, and traditional graphs—compare balance sheet layouts, emphasizing user preferences and task alignment \cite{tanlamai2011learning2}. (c) Line and bubble charts explore model behavior, variable links, and what-if scenarios to assess GDP impacts \cite{jonker2019industry}. (Reproduced from these works, © 2011–2019, with kind permission of the authors.)}
    \label{example2}
\end{figure}

Representative examples of “building basic financial literacy” include the use of a colored line graph by Vidermanova and Melusova \cite{vidermanova2015visualization} to teach mortgage structure and repayment plans (Figure \ref{example1} (a)). Two studies by Tanlamai and Soongswang \cite{tanlamai2011learning1, tanlamai2011learning2} used formats such as “graphs with numbers” (GWN), “tables with graphs” (TWG), spatial tables, and traditional graphs to present financial statements (Figures \ref{example2} (a)–(b)), focusing on cognitive load, user preferences, and task alignment. Raungpaka et al. \cite{raungpaka2013preliminary} applied a natural metaphor—the lotus flower—to visualize accounting data (Figure \ref{example1} (b)), offering a more engaging and memorable alternative to standard charts.




\begin{figure}[htbp]  
    \centering
    \includegraphics[width=0.8\columnwidth]{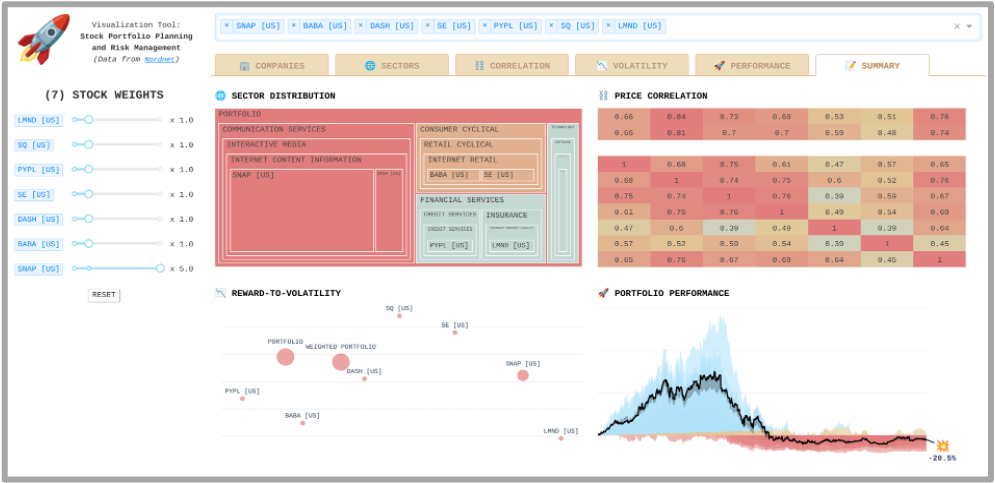} 
    \caption{Representative Example. Bar charts, treemaps, heatmaps, scatter plots, and line plots simplify financial data, revealing portfolio composition, correlations, and risk-return dynamics. (© 2024, reproduced from \cite{wang2024visualization}, with kind permission of the authors))
}
    \label{example3}
\end{figure}

\paragraph{Optimizing Investment Strategies} This task domain focuses on helping investors improve portfolio management, assess risk, and make informed decisions. Wang et al. \cite{wang2024visualization} exemplify this category by using multiple visualization techniques, as shown in Figure \ref{example3}. The bar charts display performance relative to market value, treemaps illustrate sector diversification, heatmaps reveal price correlations, scatter plots show risk-adjusted returns, and line plots track portfolio trends.

\paragraph{Decoding Financial Contexts}
This domain helps users interpret financial news and socio-economic relationships through narrative and interactive visualizations, revealing patterns and connections between financial events and broader factors. In Figure \ref{example1} (c), Chan and Qu \cite{chan2016finavistory} introduce FinaVistory, which combines stacked bar charts to show news sentiment and price trends, word clouds for key topics, and network graphs to map thematic links. A central heatmap tracks topic relevance to price changes, while customizable bubble charts let users explore user-defined variables and price dynamics.

\paragraph{Interpreting Complex Financial Systems}
This domain helps users analyze complex models and datasets to uncover patterns and understand financial dynamics. Lugmayr et al. developed ElectrAus, an immersive 3D prototype for visualizing energy and financial systems, shown in Figure \ref{example1} (d). It includes 3D scenes for seasonal electricity demand, wind turbines for renewable energy ratios, coins for daily prices, and line charts for demand, generation, and imports. Similarly, Jonker et al. use interactive line and bubble charts to explore model behavior and compare GDP growth across countries, as illustrated in Figure \ref{example2} (c). These tools support scenario analysis, highlight variable relationships, and simplify complex datasets to reveal model structures and sensitivities, enhancing decision-making.

\paragraph{Identified Gaps and Future Directions}
The four task domains align with Lusardi’s \cite{lusardi2015financial} framework, which draws on the Programme for International Student Assessment (PISA) categories: Money and transactions, Planning and managing finances, Risk and reward, and Financial landscape. Building financial literacy relates to the first two, focusing on budgeting, saving, and money management. Optimizing investment strategies aligns with Risk and reward. Decoding financial contexts connects to both Risk and reward and Financial landscape. Interpreting complex systems addresses Financial landscape by examining systemic risks and market dynamics.

This alignment reveals key gaps. Few tools target children and adolescents, despite the importance of early education in core areas. Interactive methods for teaching Risk and reward remain limited, reducing engagement. Complex domains like Decoding and Interpreting often cater to experts, leaving non-experts underserved. Educational tools rarely integrate multiple content areas, limiting their ability to convey real-world financial complexity. Future research should develop age-appropriate, inclusive, and engaging tools that incorporate Lusardi’s framework into holistic financial education.

\subsection{How were the teaching interventions evaluated and what results were obtained about their effectiveness?}

In this section, we first summarize the evaluation study designs of the 37 papers, then outline participant demographics, and finally examine key findings and statistical validity. By analyzing how visualization-based teaching interventions were evaluated and what results were reported, particularly in relation to learner understanding and experiences, we gain insights into their effectiveness, addressing our research question.

\subsubsection{Evaluation Design}
Evaluation design in this review refers to how authors assessed their studies, including methods, instruments, formats, and durations. Understanding these practices offers useful guidance for designing effective evaluations and helps readers better judge the quality and reliability of the findings.

\begin{table*}[htbp]
    \centering
    \renewcommand{\arraystretch}{1.2}
    \caption{Summary of Evaluation Design}
    \resizebox{1\textwidth}{!}{
    \begin{tabular}{>{\centering\arraybackslash}m{2cm}>{\centering\arraybackslash}m{7cm}>
    {\centering\arraybackslash}m{1.5cm}>
    {\centering\arraybackslash}m{2cm}>{\centering\arraybackslash}m{2.2cm}>
    {\centering\arraybackslash}m{1.8cm}>{\centering\arraybackslash}m{4cm}>
    {\centering\arraybackslash}m{3cm}
    }
        \hline
        \multirow{2}[2]{2cm}{\centering\arraybackslash\textbf{Paper}} & 
        \multirow{2}[2]{7cm}{\centering\arraybackslash\textbf{Instrument}} & 
        \multicolumn{3}{c}{\textbf{Format}} & 
        \multicolumn{3}{c}{\textbf{Duration}} \\ 
        \multicolumn{2}{c}{} & \rule{6.8cm}{0.4pt} & & & \rule{9.8cm}{0.4pt} \\ 
        & & \textbf{Online} 
        & \textbf{In-person} 
        & \textbf{Unspecified} 
        & \textbf{One-off} 
        & \textbf{Repeated one-off} 
        & \textbf{Longitudinal} \\ 
        \hline
        
        \cite{kothakota2020use} & Survey & \texttimes & & & \texttimes & & \\ 
        \cite{savikhin2008applied} & Test/Quiz, Survey & & & \texttimes & \texttimes & & \\ 
        \cite{rudolph2009finvis} & Questionnaire & & \texttimes & & \texttimes & & \\
        \cite{lemieux2014using} & Survey & & \texttimes & & \texttimes & & \\
        \cite{tanlamai2011learning1} & Questionnaire & \texttimes & & & & \texttimes & \\
        \cite{tanlamai2011learning2} & Questionnaire & \texttimes & & & \texttimes & & \\ 
        \cite{raungpaka2013preliminary} & Questionnaire & \texttimes & & & \texttimes & & \\ 
        \cite{stanca2020impact1} & Questionnaire, Test/Quiz & \texttimes & & & & & \texttimes \\
        \cite{cainas2021kat} & Survey & & & \texttimes & \texttimes & & \\
        \cite{tadesse2022combining} & Questionnaire & \texttimes & & & \texttimes & & \\
        \cite{savikhin2011experimental} & Questionnaire, Survey & & \texttimes & & \texttimes & & \\ 
        \cite{janvrin2014making} & Questionnaire & & & \texttimes & \texttimes & & \\
        \cite{zhang2014incorporating} & Test/Quiz, Survey & & & \texttimes & \texttimes & & \\  
        \cite{vidermanova2015visualization} & Survey & & \texttimes & & \texttimes & & \\
        \cite{mendez2015visualizing} & Test/Quiz, Interview, Observation & & \texttimes & & \texttimes & & \\ 
        \cite{kokina2017role} & Questionnaire, Test/Quiz & & & \texttimes & \texttimes & & \\
        \cite{staveley2018integrating} & Test/Quiz & & \texttimes & & \texttimes & & \\ 
        \cite{jonker2019industry} & Unspecified & & & \texttimes & & & \texttimes \\ 
        \cite{schonfeldt2020ict} & Reflection exercise & \texttimes & & & & & \texttimes \\ 
        \cite{prokofieva2020visualization} & Test/Quiz & \texttimes & & & & \texttimes & \\ 
        \cite{laplante2021incorporating} & Interview, Survey & & & \texttimes & & \texttimes &  \\
        \cite{choi2023enhancing} & Unspecified & & & \texttimes & \texttimes & & \\
        \cite{kidwell2024tableau} & Questionnaire & & & \texttimes & & \texttimes & \\ 
        \cite{stanca2020impact2} & Questionnaire & & & \texttimes & & & \texttimes \\
        \cite{schroder2022pension} & Interview & & \texttimes & & \texttimes & & \\ 
        \cite{tomasi2023role} & Survey & \texttimes & & & \texttimes & & \\ 
        \cite{hirsch2015visualisation} & Questionnaire & & \texttimes & & \texttimes & & \\
        \cite{lusardi2017visual} & Questionnaire & \texttimes & & & \texttimes & & \\
        \hline
        SUM &  & 10 & 8 & 10 & 20 & 4 & 4 \\ 

        \hline
        \textbf{Others} &  \multicolumn{7}{l}{Six Papers
        \cite{csallner2003fundexplorer, ervin2004visualizing, helweg2007business, lugmayr2019financial, varma2022web, wang2024visualization} did not mention any form of evaluation.}\\ 
        & \multicolumn{7}{l}{ One paper \cite{chan2016finavistory} reported conducting a case study.}\\
        & \multicolumn{7}{l}{Two papers \cite{ito2018text, ito2020ginn} evaluated functions by comparing them with other methods, but the evaluation did not involve users.} \\
        \hline
        
    \end{tabular}
    }
    \label{Evaluation Design}
\end{table*}

We first discuss evaluation methods. Among the 37 papers, 6 lacked evaluation \cite{csallner2003fundexplorer, ervin2004visualizing, wang2024visualization, helweg2007business, lugmayr2019financial, varma2022web}, and one used a case study \cite{chan2016finavistory}. Two papers by Ito et al. \cite{ito2018text, ito2020ginn} compared their approaches to other methods without involving users. Excluding these 9, 28 papers reported user studies. Among them, Méndez-Carbajo \cite{mendez2015visualizing} and Staveley-O'Carroll \cite{staveley2018integrating} briefly mentioned a test and an observation without detail.

We categorized the 28 papers with user studies into four groups, as shown in Table \ref{main_table_5_year_trend}. 1) Pre-test/post-test (15 papers): measured the effect of an intervention by assessing participants’ knowledge before and after. 2) Comparative testing (9 papers): compared the intervention with a control condition or alternative interventions. 3) Summative assessment (12 papers): only measured learning at the end of the study, i.e., it is only assessing what knowledge participants had at the end, but not whether new knowledge was gained. An additional 9 papers did not fit these categories and were grouped as “Others,” as defined in Table \ref{Evaluation Design}. Among the 28 papers with user studies, 19 employed statistical analysis. The distribution of evaluation types shows a clear preference for pre- and post-tests, which have become more common over the past 25 years, reflecting a focus on measuring knowledge or skill gains. Although less common, use of comparative studies is growing. 

Furthermore, Table \ref{Evaluation Design} summarizes the instruments, formats, and durations used in the evaluations and defines the “Others” category. In terms of instruments, 13 studies used self-designed questionnaires, 9 used surveys, 7 included tests or quizzes, 3 conducted interviews \cite{mendez2015visualizing, laplante2021incorporating, schroder2022pension}, one included observation \cite{mendez2015visualizing}, and one involved a reflection exercise \cite{schonfeldt2020ict}. Two studies did not specify their instruments \cite{jonker2019industry, choi2023enhancing}. Many studies combined methods. Pre- and post-test designs often included quizzes (e.g., \cite{savikhin2008applied, kokina2017role, stanca2020impact1}), while A/B testing studies relied on surveys or questionnaires (e.g., \cite{tanlamai2011learning1, hirsch2015visualisation, kothakota2020use}), though their evaluation focus varied. Most studies used self-designed instruments, with few employing standardized scales. This may indicate a lack of uniform evaluation standards, limiting comparability across studies, but also reflects the diverse contexts in which financial education research is conducted.

When examining evaluation formats, a near balance emerges between online (10 papers) and in-person (8 papers) studies. The choice of format closely aligns with data collection methods: online formats are favored for surveys and questionnaires \cite{tanlamai2011learning2, stanca2020impact1}, while in-person settings are used for interviews or observations \cite{mendez2015visualizing}, allowing for richer insights into user behavior. This suggests that while digital tools offer efficiency and scalability, in-person formats remain valuable for exploring emotional or contextual aspects. Interestingly, 10 papers did not specify the study format, suggesting room for clearer reporting in future work.

In terms of duration, most studies (20 papers) used one-off designs, collecting data at a single time point, reflecting a reliance on short-term evaluation. Four papers \cite{tanlamai2011learning1, prokofieva2020visualization, laplante2021incorporating, kidwell2024tableau} conducted repeated one-off studies, applying the same design to different groups over time, which reflects efforts to balance short- and long-term approaches. Longitudinal designs appeared in four studies, involving repeated data collection from the same participants. For example, Sch\"onfeldt and Birt \cite{schonfeldt2020ict} tracked learning over 12 weeks, and Stanca et al. \cite{stanca2020impact1, stanca2020impact2} conducted a two-year and a 14-week study. Jonker et al. \cite{jonker2019industry} reported long-term engagement without specifying duration. 

\subsubsection{Participant Demographics}

Detailed demographic information from all 37 papers and the definition of “Others” are provided in the supplementary material. Besides the 9 ``Others” in Table \ref{Evaluation Design}, 4 papers reported evaluation but lacked demographic data or discussion \cite{lemieux2014using, staveley2018integrating, jonker2019industry, cainas2021kat}.

Demographic data varied widely, making study results difficult to compare. Sample sizes spanned from very small (16 participants \cite{zhang2014incorporating}) to very large (1,797 \cite{tanlamai2011learning1}), revealing inconsistent scales across studies. Gender ratios were equally uneven: some studies achieved balance \cite{schroder2022pension, choi2023enhancing}, while others leaned towards female \cite{stanca2020impact1, stanca2020impact2} or male \cite{savikhin2011experimental, kidwell2024tableau}. Most participants were university students aged 20–25, though a few extended the age range to 18–60 \cite{kothakota2020use, choi2023enhancing}, involved younger learners aged 16–17 \cite{vidermanova2015visualization}, or included middle-aged adults aged 38–55 \cite{schroder2022pension}. The majority came from business-related backgrounds \cite{stanca2020impact1, savikhin2011experimental, laplante2021incorporating}, often as beginners \cite{schonfeldt2020ict}. Notably, only one study involved grammar school students \cite{vidermanova2015visualization}, and only two reported ethnic data \cite{kothakota2020use, lusardi2017visual}, highlighting a narrow demographic focus in most research.

Our results suggest that variations in sample size, gender, academic major, and background knowledge likely reflect participants’ academic fields or course contexts. Given significant disparities in financial literacy across ethnic groups \cite{angrisani2021racial}, the lack of ethnicity reporting may limit understanding of how these differences influence the usability or effectiveness of financial visualizations. Moreover, university-level financial literacy education dominates, which likely stems from the accessibility of university participants for academic research. Lusardi and Mitchell \cite{lusardi2009financial} emphasize that financial education should start early, while Nelson \cite{nelson2010financial} highlights the urgent need to improve financial literacy across all age groups. The lack of research focused on younger age groups, particularly those using visualizations, may be due to researchers overlooking this population or because relevant studies did not employ visualization and were therefore excluded from this review.

\subsubsection{Key Findings and Statistical Validity}
\paragraph{Evaluation Patterns and Insights}

To systematically analyze 37 papers on financial visualization tools and evaluations involving learners or users, we identified 14 core dimensions of key findings, including perceived and measured effectiveness of financial visualization tools, user experiences (e.g., enjoyment, confidence, perceived and measured performance, and engagement), and intentions for future use. For perceived and measured effectiveness, we specifically categorized evaluations into perceived effectiveness (financial literacy), measured effectiveness (financial literacy), perceived effectiveness (visualization tool), and measured effectiveness (visualization tool), distinguishing whether the focus was on improving financial literacy or assessing the visualization tool itself. Similarly, for perceived and measured performance, we distinguished between perceived performance (financial literacy), measured performance (financial literacy), perceived performance (visualization tool), and measured performance (visualization tool), based on whether the evaluations targeted the users ability to effectively use the visualization tool or their performance in financial literacy tasks. 

The evaluation results for each dimension were categorized into five groups: high (indicated positive results), neutral (reflected mixed or moderate findings), low (represented negative or minimal impact), unspecified (evaluated but no findings reported), and none (not evaluated). These classifications were based on our close reading of each paper’s reported outcomes, coding them according to stated findings in results or discussion sections.

Additionally, we annotated statistical significance by distinguishing qualitative analysis (QL) from quantitative analysis (QN), with QN further classified as statistically significant (SS) or not statistically significant (NSS). NSS results were divided into three subgroups: \( p \geq 0.05 \), no p-value reported (Np), and and quantitative analysis where no statistical testing was conducted (NST).

\begin{figure*}[htbp]
    \centering
    \includegraphics[width=\textwidth]{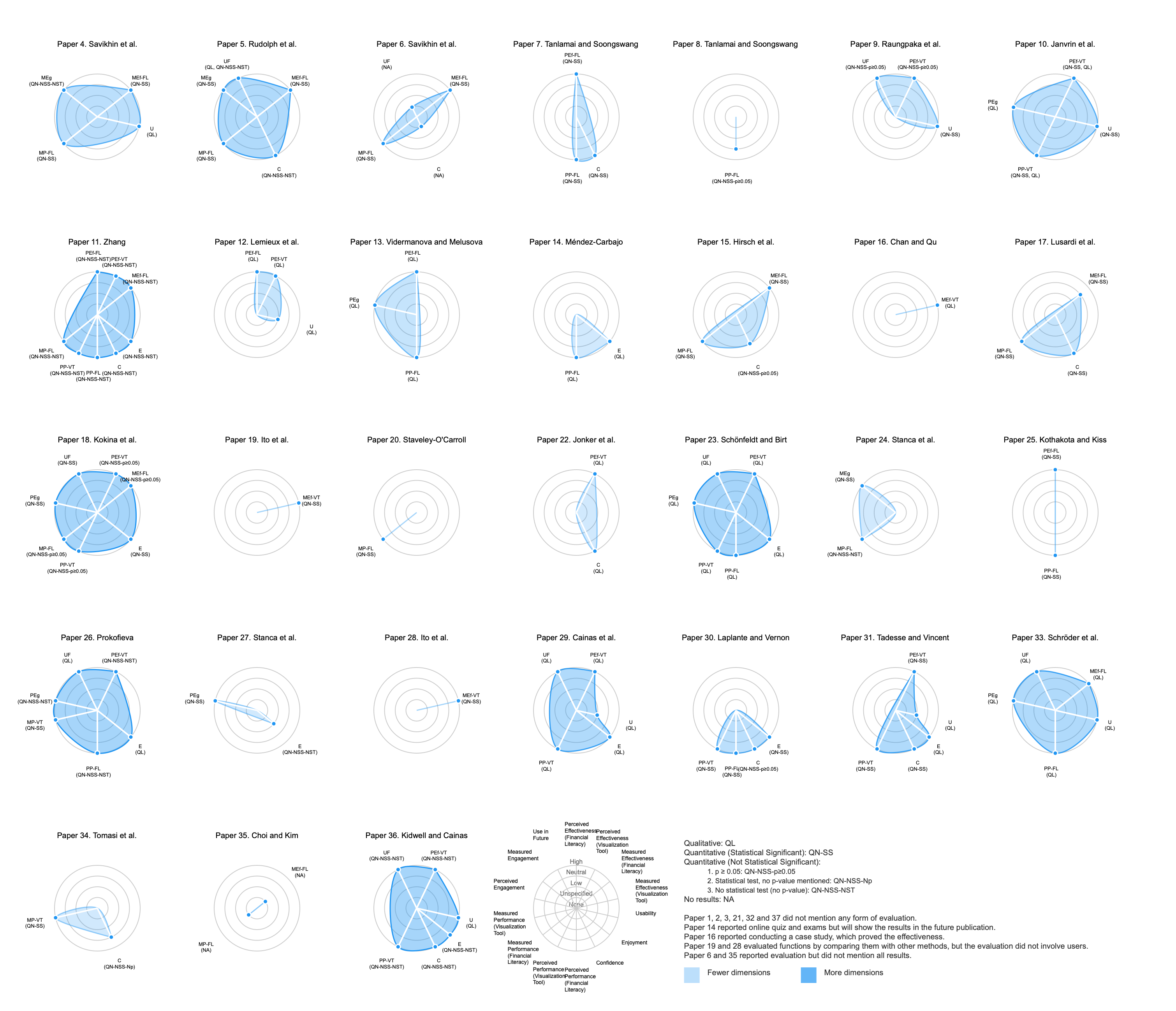}
    \caption{Small multiples radar chart of key research findings and statistical significance. Each radar chart plots a single paper across 14 evaluation dimensions, using axis labels and statistical markers in parentheses, with fill opacity indicating the breadth of evaluation scope, arranged in a grid for easy cross‐comparison.}
    \label{Small-multiples-radar-chart}
\end{figure*}

To visualize evaluation scope, key findings, and statistical significance, we use a small multiples radar chart, shown as Figure \ref{Small-multiples-radar-chart}, enabling consistent cross-comparison and pattern identification. Each paper is shown as a radar chart in a grid layout, with 14 evaluation dimensions radiating from the center. Evaluation scope ranges from none, unspecified, low, neutral, to high, with dimension labels on the axes and statistical categories in parentheses. Fill opacity reflects the number of dimensions assessed—higher opacity indicates broader scope. In the bottom right, the legend explains the encodings and statistical markers, while the annotations highlight key evaluation gaps.

Figure \ref{Small-multiples-radar-chart} reveals several noteworthy patterns in evaluation practices. First, while a few studies \cite{zhang2014incorporating, kokina2017role} evaluated over seven dimensions, indicating a broad scope, many others \cite{kokina2017role, chan2016finavistory, ito2018text, staveley2018integrating, ito2020ginn} assessed only one, reflecting narrower, more targeted strategies. Second, certain dimensions receive more consistent attention. Perceived Effectiveness (Visualization Tool), Confidence, Enjoyment, Perceived Performance (Financial Literacy), and Measured Performance (Financial Literacy) are the most frequently evaluated, each appearing in 10–11 papers, suggesting their central role. In contrast, infrequently evaluated dimensions highlight underexplored areas for future research. Third, patterns across dimensions suggest possible interrelationships: high Perceived Performance often co-occurs with high Confidence and Usability, as seen in \cite{tanlamai2011learning1, kidwell2024tableau}, raising questions about causal links and their implications for tool design.

When examining statistical significance, 13 papers rely on qualitative methods, capturing user perceptions without statistical tests. Among quantitative studies, 19 report statistically significant results (QN-SS), mostly through comparative testing. Several found their interventions more effective than control conditions, with some compared to tables \cite{rudolph2009finvis, savikhin2008applied, hirsch2015visualisation, tanlamai2011learning1} and others to text \cite{kothakota2020use, savikhin2011experimental}. Only one study \cite{lusardi2017visual} found videos most effective in improving financial literacy, while visualization tools increased the users confidence but not their literacy scores.

Additionally, several papers present quantitative findings without reaching statistical significance: 5 papers report results with \( p \geq 0.05 \) (QN-NSS-p \( \geq \) 0.05), 7 papers do not perform statistical tests or provide p-values (QN-NSS-NST), and 1 paper \cite{tomasi2023role} conducts statistical tests but does not report p-values (QN-NSS-Np). This variation highlights the need for more rigorous and transparent reporting to strengthen the reliability and comparability of findings in financial education research.

While many studies affirm the value of visualization tools in financial education, the strength and type of evidence vary considerably. Some studies focus on statistical effectiveness, with 11 showing their visualization tools are effective, 13 reporting improved student performance, 6 demonstrating superiority over alternative methods, and 14 papers noting positive learner feedback. However, 12 studies report no evidence of learning improvement, revealing inconsistencies in how effectiveness is measured. This divergence suggests that although visualization is widely seen as beneficial, empirical support remains uneven, pointing to a need for clearer evaluation frameworks.

\begin{table*}[htbp]
    \centering
    \renewcommand{\arraystretch}{1.2} 
    \caption{Collection Methods by Key Finding Dimensions}
    \resizebox{1\textwidth}{!}{
    \begin{tabular}{>{\centering\arraybackslash}m{4cm} >{\centering\arraybackslash}m{2.5cm} >{\centering\arraybackslash}m{3.5cm} >{\centering\arraybackslash}m{1.8cm} >{\centering\arraybackslash}m{2cm} >{\centering\arraybackslash}m{2.8cm} >{\centering\arraybackslash}m{3cm} >{\centering\arraybackslash}m{2.8cm} }
        \hline
        \multirow{2}[4]{*}{\textbf{Key Finding Dimension}} 
        & \multicolumn{7}{c}{\textbf{Collection Method}} 
        \\ 
        \cline{2-8}
        & \textbf{Open-ended answer} 
        & \textbf{Scale questions} 
        & \textbf{Multiple-choice} 
        & \textbf{Structured interview} 
        & \textbf{Exam performance} 
        & \textbf{Experimental measurements} 
        & \textbf{Case-based experiment}
        \\ 
        \hline
        
        \textbf{Perceived Effectiveness (Financial Literacy) } & 
        \cite{lemieux2014using}
        \cite{vidermanova2015visualization}
         & 
        \cite{tanlamai2011learning1} 
        \cite{zhang2014incorporating} & 
        \cite{kothakota2020use} & & & & \\
        \textbf{Perceived Effectiveness (Visualization Tool)} 
        & 
        \cite{lemieux2014using}
        \cite{cainas2021kat} 
        \cite{jonker2019industry}
        \cite{schonfeldt2020ict} 
        \cite{prokofieva2020visualization}
        & 
        \cite{raungpaka2013preliminary}
        \cite{tadesse2022combining}
        \cite{janvrin2014making}
        \cite{zhang2014incorporating}
        \cite{kokina2017role}
        \cite{kidwell2024tableau}
        & 
        \cite{prokofieva2020visualization} & & & & \\
        \textbf{Measured Effectiveness (Financial Literacy)} & & & \cite{lusardi2017visual} & 
        \cite{schroder2022pension} & 
        \cite{zhang2014incorporating}
        \cite{hirsch2015visualisation}
        & 
        \cite{savikhin2008applied}
        \cite{rudolph2009finvis}
        \cite{savikhin2011experimental}
        \cite{kokina2017role}
        \cite{choi2023enhancing}
        & \\
        \textbf{Measured Effectiveness (Visualization Tool)} & & &  &  &  & 
        \cite{chan2016finavistory} & 
        \cite{ito2018text}
        \cite{ito2020ginn} \\
        \textbf{Usability} & 
        \cite{savikhin2008applied}
        \cite{lemieux2014using}
        \cite{cainas2021kat}
        \cite{tadesse2022combining}
        \cite{kidwell2024tableau}
        \cite{schroder2022pension}
         & 
        \cite{raungpaka2013preliminary}
        \cite{janvrin2014making} &  &  &  &  & \\
        \textbf{Enjoyment} & 
        \cite{cainas2021kat}
        \cite{tadesse2022combining}
        \cite{mendez2015visualizing}
        \cite{schonfeldt2020ict}
        \cite{prokofieva2020visualization}
        & 
        \cite{zhang2014incorporating} 
        \cite{kokina2017role}
        \cite{laplante2021incorporating}
        \cite{kidwell2024tableau}
        \cite{stanca2020impact2} &  &  &  &  & \\
        \textbf{Confidence} & 
        \cite{jonker2019industry} 
        \cite{kidwell2024tableau} & 
        \cite{rudolph2009finvis}
        \cite{tanlamai2011learning1}
        \cite{tadesse2022combining}
        \cite{savikhin2011experimental}
        \cite{zhang2014incorporating}
        \cite{laplante2021incorporating}
        \cite{tomasi2023role} 
        \cite{hirsch2015visualisation}
        \cite{lusardi2017visual}
        &  &  &  &  & \\
        \textbf{Perceived Performance (Financial Literacy)} & 
        \cite{vidermanova2015visualization}
        \cite{mendez2015visualizing}
        \cite{schonfeldt2020ict}
        \cite{prokofieva2020visualization}
        \cite{schroder2022pension} & 
        \cite{tanlamai2011learning1}
        \cite{tanlamai2011learning2} 
        \cite{zhang2014incorporating} 
        \cite{laplante2021incorporating} & 
        \cite{kothakota2020use} &  &  &  & \\
        \textbf{Perceived Performance (Visualization Tool)} & 
        \cite{cainas2021kat}
        \cite{schonfeldt2020ict} & 
        \cite{tadesse2022combining}
        \cite{janvrin2014making}
        \cite{zhang2014incorporating}
        \cite{kokina2017role}
        \cite{laplante2021incorporating}
        \cite{kidwell2024tableau} &  &  &  &  & \\
        \textbf{Measured Performance (Financial Literacy)} & & & 
        \cite{lusardi2017visual} &  &  
        \cite{stanca2020impact1}
        \cite{zhang2014incorporating}
        \cite{staveley2018integrating} 
        \cite{hirsch2015visualisation}
         & 
        \cite{savikhin2008applied}
        \cite{rudolph2009finvis}
        \cite{savikhin2011experimental}
        \cite{kokina2017role}
        \cite{choi2023enhancing} & \\
        \textbf{Measured Performance (Visualization Tool)} & & 
        \cite{tomasi2023role} & 
        \cite{prokofieva2020visualization} &  &  &  & \\
        \textbf{Perceived Engagement} & 
        \cite{janvrin2014making}
        \cite{vidermanova2015visualization}
        \cite{schonfeldt2020ict}
        \cite{prokofieva2020visualization}
        \cite{schroder2022pension} & 
        \cite{kokina2017role} 
        \cite{stanca2020impact2} &  &  &  &  & \\
        \textbf{Measured Engagement} & & 
        \cite{stanca2020impact1} &  &  &  &  
        \cite{savikhin2008applied} 
        \cite{rudolph2009finvis} & \\
        \textbf{Use in Future} & 
        \cite{rudolph2009finvis}
        \cite{cainas2021kat}
        \cite{kokina2017role} 
        \cite{schonfeldt2020ict} 
        \cite{prokofieva2020visualization}
        \cite{schroder2022pension} & 
        \cite{raungpaka2013preliminary}
        \cite{savikhin2011experimental}
        \cite{kokina2017role}
        \cite{kidwell2024tableau} &  &  &  &  & \\
        \hline
        \textbf{Dimension SUM} & 9 & 11 & 6 & 1 & 2 & 4 & 1 \\
        \textbf{Paper SUM} & 14 & 16 & 3 & 1 & 4 & 6 & 2 \\
        \hline
    \end{tabular}
    }
    \label{Collection Methods}
\end{table*}

\paragraph{Collection Methods and Statistical Analysis}

To better understand the evaluation methods and outcomes of financial visualization tools, we analyzed the collection methods and result types for each evaluation dimension, as shown in Table \ref{Collection Methods} and Table \ref{Result Types}. Table \ref{Collection Methods} summarizes how key findings (e.g., perceived effectiveness, usability, enjoyment, confidence, performance) were collected. Table \ref{Result Types} classifies result types into post-intervention outcomes, pre-post comparisons, and cross-intervention comparisons.

\begin{table*}[htbp]
    \centering
    \renewcommand{\arraystretch}{1.2} 
    \caption{Result Types by Key Finding Dimensions}
    \resizebox{1\textwidth}{!}{
    \begin{tabular}{>{\centering\arraybackslash}m{4cm} >{\centering\arraybackslash}m{6cm} >{\centering\arraybackslash}m{5cm} >{\centering\arraybackslash}m{6cm}}
    
        \hline
        \multirow{2}{*}{\textbf{Key Finding Dimension}} 
        & \multicolumn{3}{c}{\textbf{Result Type}}  
        \\ 
        \cline{2-4} 
        & \textbf{Post-intervention} 
        & \textbf{Pre-post Intervention} 
        & \textbf{Cross-intervention} 
        \\ 
        \hline
        
        \textbf{Perceived Effectiveness (Financial Literacy)} & 
        \cite{lemieux2014using}
        \cite{zhang2014incorporating}
        \cite{vidermanova2015visualization} && 
        \cite{kothakota2020use} 
        \cite{tanlamai2011learning1}
        \\
        \textbf{Perceived Effectiveness (Visualization Tool)} & 
        \cite{lemieux2014using}
        \cite{raungpaka2013preliminary}
        \cite{cainas2021kat}
        \cite{tadesse2022combining}
        \cite{zhang2014incorporating}
        \cite{kokina2017role}
        \cite{jonker2019industry}
        \cite{schonfeldt2020ict} 
        \cite{prokofieva2020visualization}
        \cite{kidwell2024tableau} & 
        \cite{janvrin2014making} &  \\
        \textbf{Measured Effectiveness (Financial Literacy)} & 
        \cite{zhang2014incorporating} 
        \cite{choi2023enhancing} & 
        \cite{kokina2017role}
        \cite{schroder2022pension} & 
        \cite{savikhin2008applied} 
        \cite{rudolph2009finvis}
        \cite{savikhin2011experimental} 
        \cite{choi2023enhancing}
        \cite{hirsch2015visualisation}
        \cite{lusardi2017visual} \\
        \textbf{Measured Effectiveness (Visualization Tool)} & 
        \cite{ito2018text}
        \cite{ito2020ginn} 
        \cite{chan2016finavistory} & &   \\
        \textbf{Usability} &  
        \cite{savikhin2008applied}
        \cite{lemieux2014using}
        \cite{raungpaka2013preliminary}
        \cite{cainas2021kat}
        \cite{tadesse2022combining}
        \cite{janvrin2014making} 
        \cite{kidwell2024tableau}
        \cite{schroder2022pension} &  &  \\
        \textbf{Enjoyment} & 
        \cite{cainas2021kat}
        \cite{tadesse2022combining}
        \cite{zhang2014incorporating}
        \cite{mendez2015visualizing} 
        \cite{kokina2017role}
        \cite{schonfeldt2020ict} 
        \cite{prokofieva2020visualization} 
        \cite{laplante2021incorporating}
        \cite{kidwell2024tableau} 
        \cite{stanca2020impact2}&  &   \\
        \textbf{Confidence} & 
        \cite{tadesse2022combining}
        \cite{savikhin2011experimental}
        \cite{zhang2014incorporating}
        \cite{jonker2019industry}
        \cite{laplante2021incorporating}
        \cite{kidwell2024tableau} &  & 
        \cite{rudolph2009finvis} 
        \cite{tanlamai2011learning1}
        \cite{tomasi2023role}
        \cite{hirsch2015visualisation}
        \cite{lusardi2017visual} \\
        \textbf{Perceived Performance (Financial Literacy)} & 
        \cite{zhang2014incorporating}
        \cite{vidermanova2015visualization}
        \cite{mendez2015visualizing}
        \cite{schonfeldt2020ict}
        \cite{prokofieva2020visualization}
        \cite{laplante2021incorporating} & 
        \cite{schroder2022pension} & 
        \cite{kothakota2020use}
        \cite{tanlamai2011learning1}
        \cite{tanlamai2011learning2}  \\
        \textbf{Perceived Performance (Visualization Tool)} &  
        \cite{cainas2021kat}
        \cite{tadesse2022combining}
        \cite{zhang2014incorporating}
        \cite{schonfeldt2020ict}
        \cite{laplante2021incorporating}
        \cite{kidwell2024tableau} & 
        \cite{janvrin2014making} 
        \cite{kokina2017role} & \\
        \textbf{Measured Performance (Financial Literacy)} & 
        \cite{zhang2014incorporating} & 
        \cite{kokina2017role}
        \cite{staveley2018integrating}
        \cite{choi2023enhancing} & 
        \cite{savikhin2008applied}
        \cite{rudolph2009finvis}
        \cite{stanca2020impact1}
        \cite{savikhin2011experimental}
        \cite{choi2023enhancing}
        \cite{hirsch2015visualisation}
        \cite{lusardi2017visual} \\
        \textbf{Measured Performance (Visualization Tool)} & & 
        \cite{prokofieva2020visualization} & 
        \cite{tomasi2023role} \\
        \textbf{Perceived Engagement} & 
        \cite{janvrin2014making}
        \cite{vidermanova2015visualization}
        \cite{kokina2017role}
        \cite{schonfeldt2020ict}
        \cite{prokofieva2020visualization}
        \cite{stanca2020impact2} & 
        \cite{schroder2022pension} &  \\
        \textbf{Measured Engagement} & 
        \cite{savikhin2008applied} & & 
        \cite{rudolph2009finvis}
        \cite{stanca2020impact1} \\
        \textbf{Use in Future} & 
        \cite{rudolph2009finvis}
        \cite{raungpaka2013preliminary}
        \cite{cainas2021kat}
        \cite{savikhin2011experimental}
        \cite{kokina2017role}
        \cite{schonfeldt2020ict}
        \cite{prokofieva2020visualization}
        \cite{kidwell2024tableau}
        \cite{schroder2022pension} &  &   \\
        \hline 
        \textbf{Dimension SUM} & 13 & 7 & 7  \\
        \textbf{Paper SUM} & 23 & 6 &  11 \\
        \hline
    \end{tabular}
    }
    \label{Result Types}
\end{table*}

First, regarding collection methods, open-ended responses (9 dimensions, 14 papers) and scale questions (11 dimensions, 16 papers) were the most frequently used, reflecting a preference for both qualitative insights and structured quantitative evaluation of user experience. In contrast, structured interviews \cite{schroder2022pension} and case-based experiments \cite{ito2018text, ito2020ginn} appeared only in Measured Effectiveness (Financial Literacy) and Measured Effectiveness (Visualization Tool), respectively. Though used less often overall, exam performance (2 dimensions, 4 papers) and experimental measurements (4 dimensions, 6 papers) were more common in evaluating Measured Effectiveness and Measured Performance (Financial Literacy).

Second, post-intervention results (23 papers) dominated across dimensions with statistically significant findings, offering immediate evidence of impact. Pre-post comparisons were less common (6 papers), mainly in Perceived and Measured Performance \cite{janvrin2014making, kokina2017role, schroder2022pension}, yet critical for assessing knowledge retention over time. Cross-intervention results (11 papers), found in Confidence, Measured Effectiveness, and Measured Performance \cite{rudolph2009finvis, lusardi2017visual, hirsch2015visualisation}, reveal the relative advantages of different teaching tools and methods. While these approaches aim to measure real learning gains, their limited use suggests gaps in evaluating learning progression and comparative effectiveness.

In summary, while financial visualization research explores diverse evaluation methods, it pays limited attention to long-term effectiveness and underserved demographics, such as older adults and young children. Future studies should move beyond short-term outcomes to examine whether these tools support better financial decisions and long-term financial well-being, particularly for groups with the greatest need for financial literacy.

\section{DISCUSSION}

\subsection{From Regional Origins to Global Trends: The Evolution of Financial Education Visualization}

Over the past two decades, publications on financial literacy visualization have steadily increased, reflecting its growing importance. While early research was led by the United States, recent contributions from Europe, Asia, and Australia signal rising global interest. Many studies stress the value of financial education \cite{martin2007literature, greenspan2005importance, clark2003ignorance, lusardi2019financial} and its global implementation \cite{hishamudin2025intention, kumar2025global, gremi2025raising}. 

This expansion highlights both advances in interactive visualization and the demand for innovative educational approaches \cite{chen2008brief, ko2016survey}. The rise of cross-cultural research also opens opportunities to examine universal patterns and local adaptations in visualization strategies.

However, few studies evaluate the long-term effects of these tools on learning, financial behavior, or security \cite{amagir2018review}, leaving a critical gap. It remains unclear whether visualizations translate into better financial decisions or sustained improvements in financial well-being.

\subsection{Core Motivations and Guiding Questions: Shaping Financial Education Research}

A dominant motivation in financial education visualization research is the desire to enhance learners’ conceptual understanding and decision-making ability, rather than merely conveying financial facts. This educational focus has grown steadily over the past two decades, reflecting an evolving interest in deeper, more interactive forms of learning \cite{kothakota2020use}.

Although most research questions were inferred rather than explicitly stated, those that were articulated tended to focus on what we categorized as “impact evaluation” and “design strategies”, with limited attention to “application methods”.

By linking motivations to research questions, this study reveals a gap in addressing practical applications. This suggests the need for future research to explore how visualization tools are used in real educational contexts. Emphasizing this connection deepens our understanding of how motivations influence research focus and can inform the development of more effective and adaptable visualization tools.

\subsection{Barriers, Pedagogy, and Feedback: Understanding Learning Dynamics}

Our findings show that investment-related topics dominate financial education visualization research, reflecting industry demands and a focus on practical application. Most studies emphasize "learning to understand concepts," particularly "interpreting visualization," and target students developing basic professional skills. This underscores the importance of foundational knowledge, though areas such as management, marketing, and advanced financial skills remain underexplored.

Patterns between learning barriers and pedagogical methods suggest that interactive learning effectively addresses data complexity and technical challenges, aligning with previous work \cite{janvrin2014making}, while project-based learning strengthens data skills. Despite its potential, scaffolded learning is underutilized and warrants further exploration to support diverse learners, including both advanced users and young children. Future research should evaluate how pedagogical methods address specific barriers across diverse learner groups.

Feedback analysis shows a focus on self-, formative, and evaluative types, while peer and summative feedback is rare. Expanding them may enhance collaboration and instructional balance.

Overall, our analysis reveals that connecting learning barriers with specific pedagogical methods and types of feedback can inform more personalized strategies in financial education. This suggests that a better alignment between learners’ challenges and instructional design may enhance learning outcomes, especially for diverse target groups.

\subsection{Visualization Tools and Frameworks: Driving Innovation in Financial Education}

Several of the 37 selected papers highlight the widespread use of commercial tools such as Tableau, Excel, and Power BI. As supported by previous research, these tools are valued for their strong alignment with industry practices \cite{nayakrole}. Additionally, self-developed tools address specific research goals, such as illustrating complex concepts or exploring innovative designs like 3D environments and natural metaphors. The reliance on commercial tools reflects the importance of preparing learners for real-world financial and business environments, while custom tools demonstrate the potential for tailored educational approaches.

Most datasets focus on investment and accounting, using 2D visualizations and dashboards to support data exploration. Interaction modes center on exploration, manipulation, and selection, enabling efficient analysis. However, a limited application of 3D visualizations and advanced interaction techniques, such as multi-view coordination, suggests untapped potential for innovation, especially for advanced learners.

Despite the widespread use of commercial tools, few studies explain why specific tools were chosen or how they support educational goals. This gap may reflect a focus on functionality over pedagogical fit. Future research should explore tool selection rationales, optimize tools for diverse learners, and integrate them with immersive or AI-driven visualizations, reflecting similar conclusions in earlier work \cite{asif2024augmented}.. More comprehensive evaluation of interaction designs could also inform best practices in engagement and personalization.

Our findings suggest that the alignment between visualization types and interaction modes plays a critical role in shaping learning outcomes. Recognizing these patterns can guide the design of more effective financial literacy tools tailored to user needs.

\subsection{Evaluation Design and Statistical Insights: Unveiling Visualizations' Impact on Financial Learning}

Many papers lack detailed evaluation designs and statistical analysis, limiting reproducibility. Most rely on pre-tests or post-tests, summative assessments, or A/B testing, emphasizing short-term impacts. This highlights the need for more longitudinal studies on sustained learning outcomes.

Most participants were university students majoring in business or accounting, with limited diversity in age, ethnicity, educational background, and financial experience. Future research should include underserved groups, such as younger learners, older adults, and culturally diverse populations. Adaptive, AI-driven interactions may help address their varied needs, as supported by findings from previous studies \cite{rane2023education}. These technologies can tailor content delivery based on users’ prior knowledge, engagement patterns, or cognitive preferences—making them particularly suitable for learners with different cultural, age, or educational backgrounds. For example, adaptive systems could offer simplified explanations for younger users, or culturally relevant scenarios for diverse communities, improving both comprehension and motivation.

Commonly evaluated dimensions include perceived performance (visualization tools), measured performance (financial literacy), enjoyment, and confidence, while measured effectiveness (visualization tools) and user engagement remain underexplored. Expanding these areas may reveal new strategies to enhance interaction, motivation, and learning. Mixed-method approaches combining qualitative and quantitative insights remain valuable for capturing diverse perspectives.

Our findings suggest that structured evaluation and attention to participant demographics are essential for understanding the impact of financial visualizations. A comprehensive approach that integrates evaluation methods, demographic characteristics, outcome findings, and statistical validity allows for deeper insights and may address blind spots in previous research. Consistent with earlier studies \cite{tang2014effects}, our findings reveal a relationship between perceived performance and confidence, highlighting the potential of diverse participant groups to deepen the understanding of visualization's educational impact.

\section{LIMITATIONS}

This systematic review has several limitations. First, it includes only English-language studies, potentially omitting relevant research in other languages. Second, the search may have missed studies involving other forms of visualization—such as cartoons, games, AR or VR—which often use “visualization” as a keyword but do not focus on data visualization and were thus excluded. Third, our computer science backgrounds may have influenced relevance judgments despite multiple screenings. Fourth, categorization relied on human judgment, which may have led us to overestimate the role of visualization. Finally, space constraints limited discussion depth.

\section{CONCLUSIONS}

Data visualization and visual analytics translate abstract financial concepts into accessible formats, promoting understanding and practical application. They offer a strong potential to enhance financial education and literacy. Our review shows that recent research has increasingly focused on applying visualization tools within financial education, especially in higher education contexts. Most studies concentrate on investment, accounting, and risk communication, often targeting students developing basic professional knowledge. While such tools have improved conceptual understanding and learner engagement, research remains limited in addressing the needs of underrepresented groups, such as younger learners, older adults, and culturally diverse communities.

Furthermore, while many tools aim to support comprehension, relatively few studies rigorously evaluate their educational effectiveness or align interaction and visualization strategies with specific learning outcomes. Interactivity is often mentioned but seldom analyzed in depth, and visualization is primarily used for interpretive rather than predictive or behavioral learning. These gaps underscore the need for more systematic design approaches and better alignment between pedagogical goals and visual features.

In addition, the integration of artificial intelligence and machine learning offers promising opportunities to address some of these limitations. For example, conversational interfaces powered by AI could deliver personalized guidance, adapt to individual learners' needs, and support more engaging, interactive learning experiences. Similarly, machine learning algorithms could analyze user interactions and learning progress in real-time, enabling dynamic adjustments to visual content and difficulty levels. Although most of the reviewed studies did not yet incorporate such techniques, emerging trends suggest that AI-driven personalization and adaptivity will play a growing role in the next generation of visualization-based financial education tools.

For visualization researchers, this review provides a structured synthesis of age-appropriate visualization strategies across diverse financial topics, helping to inform future design decisions grounded in educational relevance. For educators, it surfaces critical instructional needs, especially for underserved groups, enabling more inclusive and responsive curriculum development. For designers and developers, it delivers practical insights into effective interface and interaction design, emphasizing the alignment between visual features and specific learning goals. Together, these contributions lay a foundation for advancing financial literacy through more thoughtful, evidence-informed, and learner-centered visualization tools.

\section*{Acknowledgments}
We would like to express our sincere gratitude to all the authors who kindly granted us permission to reuse their images in this research. We also appreciate the valuable contributions of researchers in the field of financial data visualization, whose work laid the foundation for this review. Finally, we thank the editors of this journal and the anonymous reviewers for their insightful comments and constructive feedback.

{

 }

\bibliographystyle{IEEEtran}
\bibliography{references}

\begin{thebibliography}{10}
\providecommand{\url}[1]{#1}
\csname url@samestyle\endcsname
\providecommand{\newblock}{\relax}
\providecommand{\bibinfo}[2]{#2}
\providecommand{\BIBentrySTDinterwordspacing}{\spaceskip=0pt\relax}
\providecommand{\BIBentryALTinterwordstretchfactor}{4}
\providecommand{\BIBentryALTinterwordspacing}{\spaceskip=\fontdimen2\font plus
\BIBentryALTinterwordstretchfactor\fontdimen3\font minus
  \fontdimen4\font\relax}
\providecommand{\BIBforeignlanguage}[2]{{%
\expandafter\ifx\csname l@#1\endcsname\relax
\typeout{** WARNING: IEEEtran.bst: No hyphenation pattern has been}%
\typeout{** loaded for the language `#1'. Using the pattern for}%
\typeout{** the default language instead.}%
\else
\language=\csname l@#1\endcsname
\fi
#2}}
\providecommand{\BIBdecl}{\relax}
\BIBdecl

\bibitem{servon2008consumer}
L.~J. Servon and R.~Kaestner, ``Consumer financial literacy and the impact of
  online banking on the financial behavior of lower-income bank customers,''
  \emph{Journal of consumer affairs}, vol.~42, no.~2, pp. 271--305, 2008.

\bibitem{robb2009effect}
C.~A. Robb and D.~L. Sharpe, ``Effect of personal financial knowledge on
  college students’ credit card behavior,'' \emph{Journal of financial
  counseling and planning}, vol.~20, no.~1, 2009.

\bibitem{kothakota2020use}
M.~G. Kothakota and D.~E. Kiss, ``{Use of Visualization Tools to Improve
  Financial Knowledge: An Experimental Approach},'' \emph{Journal of Financial
  Counseling and Planning}, vol.~31, no.~2, pp. 193--208, 2020.

\bibitem{tiaa2022pfin}
\BIBentryALTinterwordspacing
{TIAA Institute} and {Global Financial Literacy Excellence Center (GFLEC)},
  ``Tiaa institute-gflec personal finance index 2022,'' 2022, accessed:
  2025-01-14. [Online]. Available:
  \url{https://gflec.org/wp-content/uploads/2024/10/TIAA-Institute-GFLEC-2022-Personal-Finance-P-Fin-Index_FINAL.pdf}
\BIBentrySTDinterwordspacing

\bibitem{lusardi2007financial}
A.~Lusardi and O.~S. Mitchelli, ``Financial literacy and retirement
  preparedness: Evidence and implications for financial education: The problems
  are serious, and remedies are not simple,'' \emph{Business economics},
  vol.~42, pp. 35--44, 2007.

\bibitem{sii2024investigating}
A.~F. Sii~Niel, A.~John, and B.~Berky, ``Investigating the effectiveness of
  data visualization in communicating financial information,'' 2024.

\bibitem{zhang2020research}
Y.~Zhang, S.~Hou, H.~Liu, S.~Wang, and X.~Zhang, ``Research on application of
  data visualization in finance,'' \emph{DEStech Transactions on Engineering
  and Technology Research,(ACAAI)}, 2020.

\bibitem{padilla2018decision}
L.~M. Padilla, S.~H. Creem-Regehr, M.~Hegarty, and J.~K. Stefanucci, ``Decision
  making with visualizations: a cognitive framework across disciplines,''
  \emph{Cognitive research: principles and implications}, vol.~3, pp. 1--25,
  2018.

\bibitem{zabukovec2015impact}
A.~Zabukovec and J.~Jakli{\v{c}}, ``The impact of information visualisation on
  the quality of information in business decision-making,'' \emph{International
  Journal of Technology and Human Interaction (IJTHI)}, vol.~11, no.~2, pp.
  61--79, 2015.

\bibitem{booth2021systematic}
A.~Booth, M.-S. James, M.~Clowes, A.~Sutton \emph{et~al.}, ``Systematic
  approaches to a successful literature review,'' 2021.

\bibitem{csallner2003fundexplorer}
C.~Csallner, M.~Handte, O.~Lehmann, and J.~Stasko, ``Fundexplorer: Supporting
  the diversification of mutual fund portfolios using context treemaps,'' in
  \emph{IEEE Symposium on Information Visualization 2003 (IEEE Cat. No.
  03TH8714)}.\hskip 1em plus 0.5em minus 0.4em\relax IEEE, 2003, pp. 203--208.

\bibitem{azzam2013data}
T.~Azzam, S.~Evergreen, A.~A. Germuth, and S.~J. Kistler, ``Data visualization
  and evaluation,'' \emph{New Directions for Evaluation}, vol. 2013, no. 139,
  pp. 7--32, 2013.

\bibitem{chen2008brief}
C.-h. Chen, W.~H{\"a}rdle, A.~Unwin, and M.~Friendly, ``A brief history of data
  visualization,'' \emph{Handbook of data visualization}, pp. 15--56, 2008.

\bibitem{ervin2004visualizing}
W.~Ervin, ``Visualizing uncertainty: The graphical representation of risk in
  investor communications.'' \emph{Information Design Journal \& Document
  Design}, vol.~12, no.~3, 2004.

\bibitem{helweg2007business}
R.~Helweg-Larsen and E.~Helweg-Larsen, ``Business visualization: a new way to
  communicate financial information,'' \emph{Business Strategy Series}, vol.~8,
  no.~4, pp. 283--292, 2007.

\bibitem{savikhin2008applied}
A.~Savikhin, R.~Maciejewski, and D.~S. Ebert, ``Applied visual analytics for
  economic decision-making,'' in \emph{2008 IEEE symposium on visual analytics
  science and technology}.\hskip 1em plus 0.5em minus 0.4em\relax IEEE, 2008,
  pp. 107--114.

\bibitem{rudolph2009finvis}
S.~Rudolph, A.~Savikhin, and D.~S. Ebert, ``Finvis: Applied visual analytics
  for personal financial planning,'' in \emph{2009 IEEE symposium on visual
  analytics science and technology}.\hskip 1em plus 0.5em minus 0.4em\relax
  IEEE, 2009, pp. 195--202.

\bibitem{lemieux2014using}
V.~L. Lemieux, B.~W. Shieh, D.~Lau, S.~H. Jun, T.~Dang, J.~Chu, and G.~Tam,
  ``Using visual analytics to enhance data exploration and knowledge discovery
  in financial systemic risk analysis: The multivariate density estimator,''
  \emph{iConference 2014 Proceedings}, 2014.

\bibitem{tanlamai2011learning1}
U.~Tanlamai and O.~Soongswang, ``Learning financial reports from mixed
  symbolic-spatial graphs.'' \emph{Online Submission}, 2011.

\bibitem{tanlamai2011learning2}
------, ``Learning from balance sheet visualization.'' \emph{Online
  Submission}, 2011.

\bibitem{raungpaka2013preliminary}
V.~Raungpaka, P.~Savetpanuvong, and U.~Tanlamai, ``Preliminary results-nature
  as metaphor: Innovative visualization of accounting information with lotus
  plants,'' \emph{Journal of information technology applications \&
  management}, vol.~20, no.~2, pp. 1--14, 2013.

\bibitem{stanca2020impact1}
L.~Stanca, C.~Felea, R.~Stanca, and M.~Pintea, ``The impact of visualization
  tools on the learning engagement of accounting students,'' in
  \emph{Methodologies and Intelligent Systems for Technology Enhanced Learning,
  9th International Conference, Workshops}.\hskip 1em plus 0.5em minus
  0.4em\relax Springer, 2020, pp. 148--156.

\bibitem{cainas2021kat}
J.~M. Cainas, W.~M. Tietz, and T.~Miller-Nobles, ``Kat insurance: Data
  analytics cases for introductory accounting using excel, power bi, and/or
  tableau,'' \emph{Journal of Emerging Technologies in Accounting}, vol.~18,
  no.~1, pp. 77--85, 2021.

\bibitem{tadesse2022combining}
A.~F. Tadesse and N.~E. Vincent, ``Combining data analytics with xbrl: The
  viewdrive case,'' \emph{Issues in Accounting Education}, vol.~37, no.~1, pp.
  197--215, 2022.

\bibitem{savikhin2011experimental}
A.~Savikhin, H.~C. Lam, B.~Fisher, and D.~S. Ebert, ``An experimental study of
  financial portfolio selection with visual analytics for decision support,''
  in \emph{2011 44th Hawaii International Conference on System Sciences}.\hskip
  1em plus 0.5em minus 0.4em\relax IEEE, 2011, pp. 1--10.

\bibitem{janvrin2014making}
D.~J. Janvrin, R.~L. Raschke, and W.~N. Dilla, ``Making sense of complex data
  using interactive data visualization,'' \emph{Journal of Accounting
  Education}, vol.~32, no.~4, pp. 31--48, 2014.

\bibitem{zhang2014incorporating}
C.~Zhang, ``Incorporating powerful excel tools into finance teaching,''
  \emph{Journal of Financial Education}, pp. 87--113, 2014.

\bibitem{vidermanova2015visualization}
K.~Vidermanova and J.~Melusova, ``The visualization of the schedule of the
  mortgage loan as a tool for students’ better understanding of loans,''
  \emph{Procedia-Social and Behavioral Sciences}, vol. 186, pp. 1224--1231,
  2015.

\bibitem{mendez2015visualizing}
D.~M{\'e}ndez-Carbajo, ``Visualizing data and the online fred database,''
  \emph{The Journal of Economic Education}, vol.~46, no.~4, pp. 420--429, 2015.

\bibitem{kokina2017role}
J.~Kokina, D.~Pachamanova, and A.~Corbett, ``The role of data visualization and
  analytics in performance management: Guiding entrepreneurial growth
  decisions,'' \emph{Journal of Accounting Education}, vol.~38, pp. 50--62,
  2017.

\bibitem{ito2018text}
T.~Ito, H.~Sakaji, K.~Tsubouchi, K.~Izumi, and T.~Yamashita, ``Text-visualizing
  neural network model: understanding online financial textual data,'' in
  \emph{Advances in Knowledge Discovery and Data Mining: 22nd Pacific-Asia
  Conference, PAKDD 2018, Melbourne, VIC, Australia, June 3-6, 2018,
  Proceedings, Part III 22}.\hskip 1em plus 0.5em minus 0.4em\relax Springer,
  2018, pp. 247--259.

\bibitem{staveley2018integrating}
J.~Staveley-O'Carroll, ``Integrating graphing assignments into a money and
  banking course using fred,'' \emph{The Journal of Economic Education},
  vol.~49, no.~1, pp. 72--90, 2018.

\bibitem{lugmayr2019financial}
A.~Lugmayr, Y.~J. Lim, J.~Hollick, J.~Khuu, and F.~Chan, ``Financial data
  visualization in 3d on immersive virtual reality displays: A case-study for
  data storytelling and information visualization of financial data of
  australia’s energy sector,'' in \emph{Enterprise Applications, Markets and
  Services in the Finance Industry: 9th International Workshop, FinanceCom
  2018, Manchester, UK, June 22, 2018, Revised Papers 9}.\hskip 1em plus 0.5em
  minus 0.4em\relax Springer, 2019, pp. 118--130.

\bibitem{jonker2019industry}
D.~Jonker, R.~Brath, and S.~Langevin, ``Industry-driven visual analytics for
  understanding financial timeseries models,'' in \emph{2019 23rd International
  Conference Information Visualisation (IV)}.\hskip 1em plus 0.5em minus
  0.4em\relax IEEE, 2019, pp. 210--215.

\bibitem{schonfeldt2020ict}
N.~Sch{\"o}nfeldt and J.~L. Birt, ``Ict skill development using excel, xero,
  and tableau,'' \emph{Journal of Emerging Technologies in Accounting},
  vol.~17, no.~2, pp. 45--56, 2020.

\bibitem{prokofieva2020visualization}
M.~Prokofieva, ``Visualization of financial data in teaching financial
  accounting,'' in \emph{2020 24th International Conference Information
  Visualisation (IV)}.\hskip 1em plus 0.5em minus 0.4em\relax IEEE, 2020, pp.
  674--678.

\bibitem{ito2020ginn}
T.~Ito, H.~Sakaji, K.~Izumi, K.~Tsubouchi, and T.~Yamashita, ``Ginn: gradient
  interpretable neural networks for visualizing financial texts,''
  \emph{International Journal of Data Science and Analytics}, vol.~9, no.~4,
  pp. 431--445, 2020.

\bibitem{laplante2021incorporating}
S.~K. Laplante and M.~E. Vernon, ``Incorporating data analytics in a technical
  tax setting: A case using excel and tableau to examine a firm's schedule m-3
  and tax risk,'' \emph{Issues in Accounting Education}, vol.~36, no.~2, pp.
  129--139, 2021.

\bibitem{varma2022web}
J.~R. Varma and V.~Virmani, ``Web applications for teaching portfolio analysis
  and option pricing,'' \emph{Journal of Financial Education}, vol.~48, no.~1,
  pp. 61--78, 2022.

\bibitem{choi2023enhancing}
I.~Choi and W.~C. Kim, ``Enhancing financial literacy in south korea:
  Integrating ai and data visualization to understand financial instruments’
  interdependencies,'' \emph{Societal Impacts}, vol.~1, no. 1-2, p. 100024,
  2023.

\bibitem{kidwell2024tableau}
L.~A. Kidwell and J.~M. Cainas, ``A tableau teaching application in financial
  data analytics to state local governments: A case study on louisiana local
  government,'' \emph{Journal of Emerging Technologies in Accounting}, vol.~21,
  no.~1, pp. 167--189, 2024.

\bibitem{wang2024visualization}
J.~M. Wang, E.~B. S{\o}rensen, G.~Walsh, J.~Kusnick, and S.~J{\"a}nicke, ``A
  visualization tool for private investors: Stock portfolio planning and risk
  management,'' 2024.

\bibitem{stanca2020impact2}
L.~Stanca, C.~Felea, R.~Stanca, and M.~Pintea, ``The impact of personality,
  attitude and visual decision-making dashboard tools on the learning
  engagement of economist students,'' in \emph{Methodologies and Intelligent
  Systems for Technology Enhanced Learning, 10th International
  Conference}.\hskip 1em plus 0.5em minus 0.4em\relax Springer, 2020, pp.
  106--116.

\bibitem{schroder2022pension}
K.~Schr{\"o}der, P.~Belavadi, M.~Ziefle, and A.~Calero~Valdez, ``The pension
  story-data-driven storytelling with pension data,'' in \emph{International
  Conference on Human-Computer Interaction}.\hskip 1em plus 0.5em minus
  0.4em\relax Springer, 2022, pp. 404--415.

\bibitem{tomasi2023role}
S.~D. Tomasi, J.~Liu, F.~Cheng, and C.~Han, ``The role of individual
  characteristics: How thinking style and domain expertise affect performances
  on visualization,'' \emph{Information visualization}, vol.~22, no.~3, pp.
  265--276, 2023.

\bibitem{hirsch2015visualisation}
B.~Hirsch, A.~Seubert, and M.~Sohn, ``Visualisation of data in management
  accounting reports: How supplementary graphs improve every-day management
  judgments,'' \emph{Journal of Applied Accounting Research}, vol.~16, no.~2,
  pp. 221--239, 2015.

\bibitem{lusardi2017visual}
A.~Lusardi, A.~Samek, A.~Kapteyn, L.~Glinert, A.~Hung, and A.~Heinberg,
  ``Visual tools and narratives: New ways to improve financial literacy,''
  \emph{Journal of Pension Economics \& Finance}, vol.~16, no.~3, pp. 297--323,
  2017.

\bibitem{chan2016finavistory}
Y.-Y. Chan and H.~Qu, ``Finavistory: Using narrative visualization to explain
  social and economic relationships in financial news,'' in \emph{2016
  International Conference on Big Data and Smart Computing (BigComp)}.\hskip
  1em plus 0.5em minus 0.4em\relax IEEE, 2016, pp. 32--39.

\bibitem{lusardi2015financial}
A.~Lusardi, ``Financial literacy skills for the 21st century: Evidence from
  pisa,'' \emph{Journal of consumer affairs}, vol.~49, no.~3, pp. 639--659,
  2015.

\bibitem{angrisani2021racial}
M.~Angrisani, S.~Barrera, L.~R. Blanco, and S.~Contreras, ``The racial/ethnic
  gap in financial literacy in the population and by income,''
  \emph{Contemporary Economic Policy}, vol.~39, no.~3, pp. 524--536, 2021.

\bibitem{lusardi2009financial}
A.~Lusardi and O.~S. Mitchell, ``Financial literacy: Evidence and implications
  for financial education,'' \emph{Trends and issues}, pp. 1--10, 2009.

\bibitem{nelson2010financial}
C.~Nelson, ``Financial education for all ages,'' \emph{Innovations}, vol.~5,
  no.~2, pp. 83--86, 2010.

\bibitem{martin2007literature}
M.~Martin, ``A literature review on the effectiveness of financial education,''
  2007.

\bibitem{greenspan2005importance}
A.~Greenspan, ``The importance of financial education today,'' \emph{Social
  education}, vol.~69, no.~2, pp. 64--66, 2005.

\bibitem{clark2003ignorance}
R.~L. Clark and M.~B. d’Ambrosio, ``Ignorance is not bliss: The importance of
  financial education,'' \emph{TIAA-CREF Research Dialogue}, vol.~78, no.~1,
  2003.

\bibitem{lusardi2019financial}
A.~Lusardi, ``Financial literacy and the need for financial education: evidence
  and implications,'' \emph{Swiss journal of economics and statistics}, vol.
  155, no.~1, pp. 1--8, 2019.

\bibitem{hishamudin2025intention}
M.~Z. Hishamudin, N.~S. Kamarudin, N.~A. Hadi, and A.~Ahmad, ``Intention to use
  digital platforms for islamic financial education in malaysia: Structural
  equation model,'' \emph{Journal of Advanced Research in Applied Sciences and
  Engineering Technology}, vol.~49, no.~1, pp. 298--311, 2025.

\bibitem{kumar2025global}
R.~Kumar, ``Global trends and research patterns in financial literacy and
  behavior: A bibliometric analysis,'' \emph{Management Science Advances},
  vol.~2, no.~1, pp. 1--18, 2025.

\bibitem{gremi2025raising}
E.~Gremi, S.~{\c{C}}erri, M.~Morina, M.~Kotollaku, M.~{\c{C}}ela, and
  A.~Durmishi, ``Raising awareness of financial education among youth in the
  elbasan region, albania,'' \emph{Multidisciplinary Reviews}, vol.~8, no.~4,
  pp. 2\,025\,130--2\,025\,130, 2025.

\bibitem{ko2016survey}
S.~Ko, I.~Cho, S.~Afzal, C.~Yau, J.~Chae, A.~Malik, K.~Beck, Y.~Jang,
  W.~Ribarsky, and D.~S. Ebert, ``A survey on visual analysis approaches for
  financial data,'' in \emph{Computer Graphics Forum}, vol.~35, no.~3.\hskip
  1em plus 0.5em minus 0.4em\relax Wiley Online Library, 2016, pp. 599--617.

\bibitem{amagir2018review}
A.~Amagir, W.~Groot, H.~Maassen van~den Brink, and A.~Wilschut, ``A review of
  financial-literacy education programs for children and adolescents,''
  \emph{Citizenship, Social and Economics Education}, vol.~17, no.~1, pp.
  56--80, 2018.

\bibitem{nayakrole}
S.~Nayak, ``The role of data visualization tools in financial decision-making:
  A comparative analysis of tableau, power bi, and ssrs.''

\bibitem{asif2024augmented}
M.~Asif, A.~Mondal, S.~Soumil, A.~Das, and P.~Sahoo, ``Augmented reality and
  virtual reality in education: A transformative journey into immersive
  learning environments,'' \emph{Advances in Computational Solutions}, p. 185,
  2024.

\bibitem{rane2023education}
N.~Rane, S.~Choudhary, and J.~Rane, ``Education 4.0 and 5.0: Integrating
  artificial intelligence (ai) for personalized and adaptive learning,''
  \emph{Available at SSRN 4638365}, 2023.

\bibitem{tang2014effects}
F.~Tang, T.~J. Hess, J.~S. Valacich, and J.~T. Sweeney, ``The effects of
  visualization and interactivity on calibration in financial
  decision-making,'' \emph{Behavioral Research in Accounting}, vol.~26, no.~1,
  pp. 25--58, 2014.

\end{thebibliography}


\begin{thebibliography}{1}
\bibliographystyle{IEEEtran}

\bibitem{ref1}
{\it{Mathematics Into Type}}. American Mathematical Society. [Online]. Available: https://www.ams.org/arc/styleguide/mit-2.pdf

\bibitem{ref2}
T. W. Chaundy, P. R. Barrett and C. Batey, {\it{The Printing of Mathematics}}. London, U.K., Oxford Univ. Press, 1954.

\bibitem{ref3}
F. Mittelbach and M. Goossens, {\it{The \LaTeX Companion}}, 2nd ed. Boston, MA, USA: Pearson, 2004.

\bibitem{ref4}
G. Gr\"atzer, {\it{More Math Into LaTeX}}, New York, NY, USA: Springer, 2007.

\bibitem{ref5}M. Letourneau and J. W. Sharp, {\it{AMS-StyleGuide-online.pdf,}} American Mathematical Society, Providence, RI, USA, [Online]. Available: http://www.ams.org/arc/styleguide/index.html

\bibitem{ref6}
H. Sira-Ramirez, ``On the sliding mode control of nonlinear systems,'' \textit{Syst. Control Lett.}, vol. 19, pp. 303--312, 1992.

\bibitem{ref7}
A. Levant, ``Exact differentiation of signals with unbounded higher derivatives,''  in \textit{Proc. 45th IEEE Conf. Decis.
Control}, San Diego, CA, USA, 2006, pp. 5585--5590. DOI: 10.1109/CDC.2006.377165.

\bibitem{ref8}
M. Fliess, C. Join, and H. Sira-Ramirez, ``Non-linear estimation is easy,'' \textit{Int. J. Model., Ident. Control}, vol. 4, no. 1, pp. 12--27, 2008.

\bibitem{ref9}
R. Ortega, A. Astolfi, G. Bastin, and H. Rodriguez, ``Stabilization of food-chain systems using a port-controlled Hamiltonian description,'' in \textit{Proc. Amer. Control Conf.}, Chicago, IL, USA,
2000, pp. 2245--2249.


\end{thebibliography}

\newpage

\begin{IEEEbiography}
[{\includegraphics[width=1in,height=1.25in,clip,keepaspectratio]{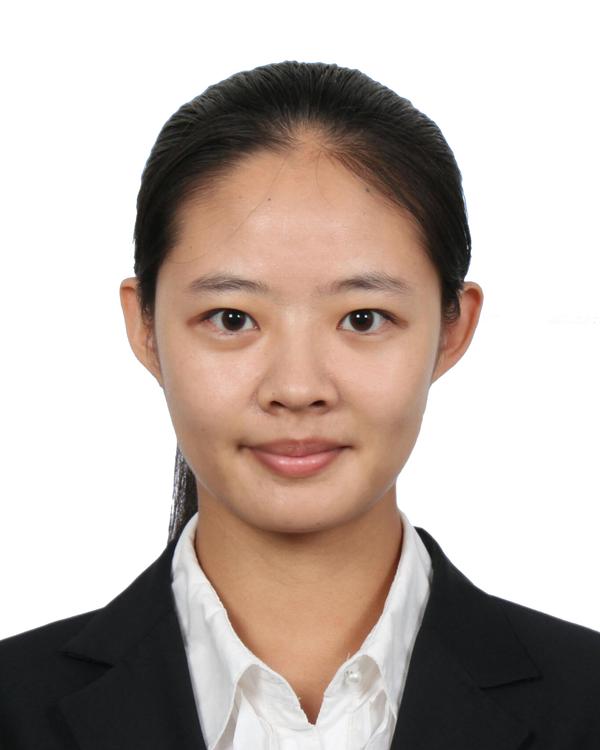}}]{Meng Du} 
is a Ph.D. candidate in computer science at the University of Auckland. She received her first Ph.D. in Forestry Equipment and Informatization from Beijing Forestry University in 2015. She later held postdoctoral research positions at the University of California, Davis, and Peking University. Her research spans computer graphics, digital media, data visualization, and visual analytics, with a particular interest in enhancing financial literacy through data visualization.
\end{IEEEbiography}

\begin{IEEEbiography}
[{\includegraphics[width=1in,height=1.25in,clip,keepaspectratio]{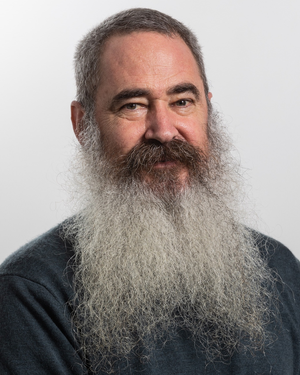}}]{Robert Amor}
is a professor in of computer science in the University of Auckland. He is also a leading researcher in Construction Informatics at the University of Auckland. His work focuses on applying computer science techniques—such as BIM, process modeling, and compliance checking—to the architecture, engineering, and construction industries. He has coordinated the international CIB W78 working group on IT in Construction since 2003 and serves as Editor-in-Chief of the Journal of Information Technologies in Construction. With over \$18 million in research funding secured, he has been instrumental in advancing interoperability and integrated design solutions within the built environment.
\end{IEEEbiography}

\begin{IEEEbiography}
[{\includegraphics[width=1in,height=1.25in,clip,keepaspectratio]{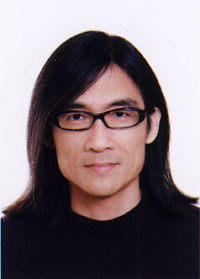}}]{Kwan-Liu Ma}
is a distinguished professor of
computer science at the University of California,
Davis. His research is in the intersection of data
visualization, computer graphics, human-computer
interaction, and high performance computing. For
his significant research accomplishments, Ma re-
ceived several recognitions including being elected
as IEEE Fellow in 2012 and ACM Fellow in 2024,
recipient of the IEEE VGTC Visualization Technical
Achievement Award in 2013, and inducted to IEEE
Visualization Academy in 2019.
\end{IEEEbiography}

\begin{IEEEbiography}
[{\includegraphics[width=1in,height=1.25in,clip,keepaspectratio]{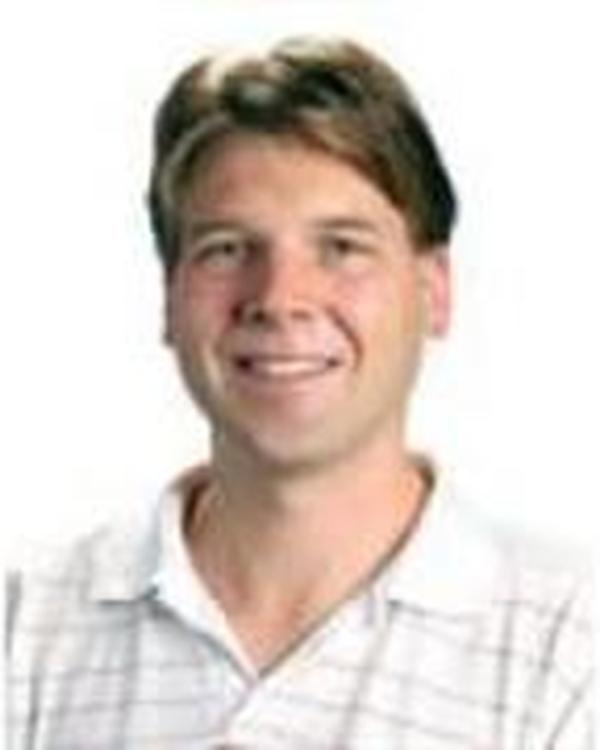}}]{Burkhard C. Wünsche}
is a senior lecturer of computer science in the University of Auckland. He is also a leading researcher in visual computing who focuses on solving real-world problems using visual representations. His research interests includes visual computing (Computer Graphics, Computer Vision, HCI, AR/VR), serious games, and innovative education applications and health interventions. Burkhard Wünsche has worked on numerous projects resulting in both high impact publications and commercial products. He led the development of ClickWorld, a novel technology making the creation of 3D models as easy as taking photos. His team won the SPARK Entrepreneurship competition and has received seed funding and support from the IceHouse business incubator.
\end{IEEEbiography}

\vspace{11pt}

\vfill

\end{document}


\onecolumn  
\appendix[Supplementary Table]

Table \ref{Participant Demographics} presents additional data on participant demographics used in the main analysis.

\renewcommand{\thetable}{A\arabic{table}}   
\setcounter{table}{0}                       

 \begin{table}[h]
    \centering
    \renewcommand{\arraystretch}{3.1}
    \caption{Summary of Participant Demographics in Evaluation Research}
    \resizebox{1\textwidth}{!}{
    \rowcolors{2}{gray!20}{white}
    \begin{tabular}{>{\centering\arraybackslash}m{1cm}>{\raggedright\arraybackslash}m{4cm}>
    {\raggedright\arraybackslash}m{4cm}>
    {\raggedright\arraybackslash}m{4cm}>{\raggedright\arraybackslash}m{6cm}>
    {\raggedright\arraybackslash}m{6cm}>{\raggedright\arraybackslash}m{6cm}>
    {\raggedright\arraybackslash}m{4cm}
    }
        \hline
        \textbf{Paper} & \textbf{Number} & \textbf{Gender} & \textbf{Age} & \textbf{Education} & \textbf{Major / Course} & \textbf{Knowledge / Experience} & \textbf{Ethnicity} \\ 
        \hline
        
        \cite{kothakota2020use} & 1,797 & $44.91\%$ M, $53.31\%$ F & 18-60+ & High school dropout: $2.67\%$, High school grad: $10.35\%$, Some college: $29.60\%$, College grad: $26.54\%$, Some graduate: $8.51\%$, Graduate school: $16.53\%$, Professional degree (JD, MD) or PhD: $5.79\%$. & & Have you taken one or more financial classes in the past year? Yes: $93.32\%$, No: $6.68\%$. & Asian/Pacific Islander ($4.23\%$), Black ($6.91\%$), Hispanic ($5.01\%$), White ($79.78\%$), Others ($4.07\%$) \\ 
        \cite{savikhin2008applied} & 56 & & & Undergraduate & & Inexperienced &\\ 
        \cite{rudolph2009finvis} & 26 & $60\%$ M, $40\%$ F & 18-23, average 21 & &  Business, engineering, information technology, science, health sciences. & &\\ 
        \cite{tanlamai2011learning1} & 828 (163 complete, 129 incomplete, 523 no response) & & Average 27.7 & University students: $59.5\%$ Regular, $40.5\%$ Executive & 25.5\% Accounting Information Systems (AIS), 61.1\% Management Information Systems (MIS), and 12.2\% Statistical Information Systems (SIT). & Average 1.96 years of experience with financial reports; backgrounds in engineering, accounting, and statistics. &\\ 
        \cite{tanlamai2011learning2} & $104$ students (56 executive, 48 regular; 62 completed) & $47.8\%$ M, $52.2\%$ F & & Master of Science & $69.9\%$ Management Information Systems (MIS), $23.0\%$ Accounting Information Systems (AIS), and $7.1\%$ Statistical Information Systems (SIT). & Executive class: at least three years of work experience; Regular class: some had work experience. Overall, $82.3\%$ had work experience. Both groups were familiar with balance sheets and accounting/finance courses. &\\ 
        \cite{raungpaka2013preliminary} & 303 (220 completed) & & 18-60, average 24.4 & $42.9\%$ Bachelor's in Accounting, $5.28\%$ Master's in Accounting, $24.4\%$ Non-Accounting (science, engineering, IT, economics) & 146 in accounting,  others in science, engineering, IT, economics. & Average 2.49 years of financial statement experience; $92.7\%$ had taken a basic accounting class; $90.5\%$ had taken a financial accounting course. &\\
        \cite{stanca2020impact1} & 743 & $24.7\%$ M, $75.3\%$ F & & & IT, Economics  & $56.2\%$ had intermediate knowledge of IT and economics, $43.8\%$ had advanced knowledge; $57.6\%$ had intermediate computer literacy, $41.5\%$ had higher intermediate literacy. &\\ 
        \cite{tadesse2022combining} & 45 & & & Graduate & Accounting & Some financial knowledge (implied) & \\ 
        \cite{savikhin2011experimental} & & $67\%$ M, $33\%$ F & 20-33, average 23 & Graduate $15\%$, Undergraduate $85\%$  & $13\%$ Arts, $25\%$ Business/Economics, $26\%$ Engineering, $13\%$ IT, $3\%$ Medicine/Nursing/Health Sciences, $20\%$ Other Sciences. &  Participants will soon enter the job market and make first-time investment decisions. &\\
        \cite{janvrin2014making} & 126 (34 field-test, 92 assignment) & & Undergraduate average 20.5, Advanced AIS participants average 23.5 & University-level (Introductory, Upper-level, Advanced) & Advanced Accounting Information Systems (AIS) course, upper-level managerial accounting course and an introductory AIS course. & Basic competency in spreadsheet graphing functions; some prior experience with financial software and data analysis.&\\ 
        \cite{zhang2014incorporating} & 16 & & & & Excel-integrated Financial Analysis and Strategy course & &\\ 
        \cite{vidermanova2015visualization} & 33 & & 16-17 & High School & Grammar School & &\\ 
        \cite{mendez2015visualizing} & & & & & Intermediate-level financial economics course & Completed introductory economics course. &\\ 
        \cite{kokina2017role} & 72 undergraduate (pre-case), 59 undergraduate (post-case), 22 graduate & Pre-case: $58.33\%$ M, $41.67\%$ F; Post-case: $50.85\%$ M, $49.15\%$ F & Undergraduate: 19 & Undergraduate ($90.28\%$ sophomore, $8.33\%$ junior, $1.39\%$ senior), Graduate & Managerial Accounting Course & MS Excel: $5.56\%$ none, $16.66\%$ beginner, $62.50\%$ intermediate, $15.28\%$ advanced. Tableau/Advanced visualization: $58.33\%$ none, $27.78\%$ beginner, $12.50\%$ intermediate, $1.39\%$ advanced. Advanced analytics (e.g., R, SPSS): $13.89\%$ none, $55.56\%$ beginner, $26.39\%$ intermediate, $4.16\%$ advanced. &\\
        \cite{schonfeldt2020ict} & $ \sim 100 $ & & & Masters & Accounting & $64.7\%$ had no or basic knowledge of Excel; $93.75\%$ had no or basic knowledge of accounting software; $100\%$ had no experience with Tableau. &\\ 
        \cite{prokofieva2020visualization} & & & & & & $30\%$ had experience in accounting firms, $40\%$ had prior experience with data analytics (including Excel), $15\%$ had experience with Tableau, $80\%$ had strong interests in data analytics and visualization. &\\ 
        \cite{laplante2021incorporating} & $ \sim 90 $ & & & $30\%$ Undergraduate, $70\%$ Master's & $90\%$ Accounting, $10\%$ Other disciplines & $40\%$ had prior exposure to Tableau, most had some Excel exposure, and few had expertise. A little over half of the students intended to pursue careers outside of tax. &\\ 
        \cite{choi2023enhancing} & 200 & Balanced representation & 18-79 & Varying levels of education & & &\\ 
        \cite{kidwell2024tableau} & 114 & $67\%$ M, $33\%$ F & 20-33, average 23 & $20\%$ Senior, $80\%$ Masters & $80\%$ Data analytics, $20\%$ Auditing & Average accounting work experience: 7 months.  Prior knowledge of big data and analytics (mean = 2.56), prior knowledge of Tableau (mean = 2.55), interest in big data and analytics (mean = 3.30). &\\
        \cite{stanca2020impact2} & & $16\%$ M, $84\%$ F & $98\%$ 18-25, $2\%$ 25-30 & Undergraduate & Economics & &\\ 
        \cite{schroder2022pension} & 10 & 6 M, 4 F & 38-55 & & & Most participants had little to no knowledge of pensions. Few had contacted their pension fund, and about half had consulted general information websites. &\\ 
        \cite{tomasi2023role} & 227 & & 18–54+ & Graduate, Undergraduate & Business & Measured by a financial knowledge test, scores ranged from 0 to 5, but does not give detailed results. &\\ 
        \cite{hirsch2015visualisation} & 120 (55 students, 65 managers) & 
        Student: $78\%$ M, $22\%$ F Manager: $72\%$ M, $28\%$ F & Students average = 24.3, Managers average = 29.7 & & & Students/Managers 4 years average work experience. &\\ 
        \cite{lusardi2017visual} & 892 & $35\text{--}45\%$ M, $55\text{--}65\%$ F & Min 18, average 49.5 & The median highest educational attainment level was ``some college, no degree.'' & & & Caucasian ($79.82\%$), African-American ($10.99\%$), Hispanic ($14.80\%$), Others ($9.19\%$)\\ 
        \hline
        \rowcolor{white} 
        \textbf{Others} 
        & \multicolumn{7}{l}{ Six papers \cite{csallner2003fundexplorer, ervin2004visualizing, helweg2007business, lugmayr2019financial, varma2022web, wang2024visualization} did not mention any form of evaluation.}\\ 
        \rowcolor{white} 
        & \multicolumn{7}{l}{One paper \cite{chan2016finavistory} reported conducting a case study.}\\
        \rowcolor{white} 
        & \multicolumn{7}{l}{Two papers \cite{ito2018text, ito2020ginn} evaluated functions by comparing them with other methods, but the evaluation did not involve users.} \\
        \rowcolor{white} 
        & \multicolumn{7}{l}{Four papers \cite{lemieux2014using, staveley2018integrating, jonker2019industry, cainas2021kat} reported evaluation but did not mention demographics.}\\
        \hline

    \end{tabular}
    }
    \label{Participant Demographics}
\end{table}

\bibliographystyle{IEEEtran}